# Respiratory Motion Management in Abdominal MRI: Revisiting the Gap Between Technical Advances and Clinical Translation

**Li Feng, PhD, and Hersh Chandarana, MD**

Center for Advanced Imaging Innovation and Research (CAI²R)
Department of Radiology
New York University Grossman School of Medicine, New York, NY, United States

**Running Head:** Motion Management in Abdominal MRI

Address correspondence to:
    Li Feng, PhD
    Associate Professor of Radiology
    Director of Rapid Imaging
    Center for Advanced Imaging Innovation and Research (CAI²R)
    New York University Grossman School of Medicine
    Email: Li.Feng@nyulangone.org

## Abstract

The inherently slow acquisition speed of MRI makes abdominal imaging highly sensitive to respiratory motion artifacts. Since the early days of MRI, the development of respiratory motion compensation has been an active research topic, and this field has achieved substantial technical progress. Despite these advances, majority of these techniques are not clinically available, and motion management methods used in clinical abdominal MRI today have changed little over the past decades. This observation is striking and points to a significant gap between technical innovation and clinical translation in this area.

This review is motivated by this question: why have so many motion management techniques not been adopted into routine clinical workflows? Unlike conventional survey-style reviews that focus on summarizing emerging methods, this article takes a different, and perhaps opposite, perspective to investigate why those technologically sophisticated innovations are misaligned with practical clinical needs. Specifically, we discuss the barriers behind the gap between research advances and clinical practice, clarify the clinical requirements for effective respiratory motion management in abdominal MRI, and highlight research directions with stronger relevance to routine workflows.

The review begins with an overview of the clinical impact of respiratory motion in abdominal MRI, followed by a discussion of standard abdominal MRI sequences and their motion sensitivity. We then summarize current clinical strategies and advanced approaches, along with the barriers that hinder their clinical adoption. The article concludes with future directions and broader lessons learned from this translational gap, with the goal of guiding future developments towards improved clinical integration.

# 1. Introduction

The sequential nature of data acquisition in MRI leads to relatively slow imaging speed compared to many other modalities, which in turn increases the sensitivity to motion artifacts. Respiratory motion therefore becomes a major source of image degradation in body MRI, particularly for abdominal organs (1–3). To mitigate these motion effects, routine abdominal MRI exams in current clinical practice typically require a series of breath-holds, with each lasting about 10-15 seconds.

A practical challenge of breath-hold imaging is that many patients, especially pediatric, elderly, or sick patients, have limited capability to hold even short breath-holds. When breath-holds fail, technologists may need to repeat acquisitions until diagnostic image quality is achieved, or switch to alternative motion mitigation strategies that often require additional scan time. In applications such as dynamic contrast-enhanced MRI (DCE-MRI) where contrast injection is usually performed only once, failed breath-holding during a rapid enhancement phase (e.g., the arterial phase) could result in motion-corrupted images that cannot be reacquired. In such cases, radiologists may have to interpret images with motion artifacts.

These limitations have motivated substantial efforts since the early days of MRI to develop approaches that allow patients to breathe normally in abdominal MRI (1–3). For example, a PubMed search using the keyword "free-breathing MRI" shows a rapid growth in related publications over the past decades, which reflects continued interest in this area. Many studies have indeed demonstrated promising solutions in controlled settings and proof-of-concept experiments. However, despite decades of technical innovation, routine abdominal MRI protocols have seen little change in motion management over the past two decades, and breath-holding remains the dominant strategy for avoiding respiratory motion artifacts. This creates a "valley of death", as illustrated in **Figure 1**, indicating a significant gap between the technical advances in respiratory motion management and their actual clinical translation and impact.

This review is motivated by a central question: why have so many motion management techniques not been adopted into routine clinical workflows? Unlike conventional survey-style reviews that primarily summarize emerging techniques, this article takes a different, and perhaps opposite, perspective to investigate why many technologically sophisticated innovations are not clinically available and are misaligned with practical needs. Specifically, we discuss the barriers behind the gap between research advances and routine clinical practice, clarify the clinical requirements for effective respiratory motion management in abdominal MRI, and highlight research directions with stronger clinical relevance in routine workflows.

This article begins with an overview of the clinical impact of respiratory motion in abdominal MRI, followed by a description of the standard MRI sequences used in routine abdominal imaging and their motion sensitivity. These sections first outline the practical problems that motion management techniques should address. We then summarize the motion management approaches used in current clinical practice and describe advanced strategies with translational potential, along with barriers that currently hinder their clinical adoption. The review concludes with a discussion of future directions in this area and the broader lessons learned from the translational gap in respiratory motion management, with the goal of guiding future developments towards improved clinical integration.

## 2. Challenges of Respiratory Motion in the Abdomen

Respiratory motion in the abdomen mainly arises from diaphragm displacement during breathing (4–6). During inspiration, diaphragm contraction moves different abdominal organs downward by several centimeters, and these organs move upward during expiration. Importantly, respiratory-induced displacement is not spatially uniform, and different organs can deform to different extents during a respiratory cycle (6). Respiratory motion occurs along both the superior-inferior (head-to-foot) and anterior-posterior directions, and displacement is typically much larger in the superior-inferior direction.

Another important feature of respiratory motion is the substantial variability in breathing patterns, both between patients and across respiratory cycles within the same patient, and baseline drift often occurs during the MRI exam. These factors together make respiratory motion often irregular and non-periodic, with spatial variation even within the same organ (6). As a result, rigid motion models are insufficient to account for the full complexity of abdominal motion, and this increases the difficulty of motion management in abdominal MRI and limits the robustness and generalizability of many motion compensation techniques.

## 3. Standard Abdominal MRI Sequences

To understand the challenges of respiratory motion management, it is important to first review the sequences routinely used in abdominal MRI, as each of them is affected by respiratory motion in different ways and no single motion management technique can serve as a universal solution. While clinical protocols vary, most routine abdominal MRI sequences can be grouped into three major categories: (1) T2-weighted imaging using 2D or 3D fast spin-echo (FSE), (2) T1-weighted imaging using single-echo or multi-echo 3D gradient-echo (GRE), and (3) 2D diffusion-weighted imaging (DWI) typically implemented using 2D echo planar imaging (EPI). This section summarizes these standard clinical sequences followed by a sequence-specific discussion of motion sensitivity.

### 3.1 T2-Weighted Imaging

Three major types of T2-weighted sequence are commonly used in clinical abdominal MRI. The first one is 2D single-shot FSE (SSFSE), which acquires each slice within a single long echo train. SSFSE is typically implemented with partial Fourier acquisition to shorten the effective echo train length, reduce scan time, and mitigate T2-related spatial blurring (7). SSFSE is widely used to provide a rapid overview of abdominal structures and for applications requiring strong T2-weighting, such as pancreatic imaging and evaluation of fluid-rich tissues. Because each slice is acquired in a single echo train (~500-800ms), SSFSE is relatively tolerant of intraframe respiratory motion and can be performed during free-breathing or breath-holding depending on clinical needs. Breath-holding is often preferred, as it gives consistent organ boundaries across slices (**Figure 2**), which is particularly important for imaging small structures such as the pancreas. Breath-hold SSFSE imaging is commonly divided into multiple breath-holds (concatenations) to cover the full imaging volume, with each concatenation acquiring a subset of slices. However, even with this approach, mild slice-to-slice inconsistency can still occur because different slice groups correspond to slightly different respiratory states over different breath-holds (see **Figure 2**), although this is often clinically acceptable. Despite its motion robustness, SSFSE is limited by spatial blurring due to long echo trains, which in turn limits its achievable spatial resolution.

The second type is multi-shot 2D FSE, which acquires each slice over multiple echo trains (also referred to as shots). Compared to SSFSE, multi-shot 2D FSE enables higher spatial resolution and reduced T2 blurring at the cost of longer scan time. Because data from different shots need to be combined into a single image, consistent respiratory state across shots is important. Therefore, multi-shot 2D FSE typically requires reliable breath-holding or other effective respiratory management strategies (see Section 4). As with SSFSE, data acquisition in multi-shot 2D FSE can be divided into multiple concatenations to shorten each breath-hold, which may result in mild slice-to-slice inconsistency, although this is generally less a concern than intraframe artifacts for this sequence.

The third type is 3D FSE, which is most commonly used for MR cholangiopancreatography (MRCP) in clinical practice (8, 9). This sequence allows extended echo trains to achieve strong T2-weighting for imaging fluid-filled structures such as the biliary and pancreatic ducts. The main advantage of 3D FSE is the nearly isotropic resolution, which allows for multiplanar reformation for improved visualization of the small ductal anatomy. However, 3D FSE requires much longer scan time than 2D FSE and is therefore highly sensitive to respiratory motion, making effective motion compensation essential for routine clinical use.

### 3.2 T1-Weighted Imaging

T1-weighted imaging in abdominal MRI is typically performed using 3D single-echo or multi-echo GRE sequences for three major applications. First, standard fat-suppressed, single-echo 3D GRE is used for pre- and post-contrast imaging, where images are acquired before and following the administration of a gadolinium-based contrast agent at multiple enhancement phases, typically over several breath-holds. This multi-phase acquisition, sometimes referred to as DCE-MRI (without pharmacokinetic modeling), is a cornerstone of routine abdominal MRI for detection, delineation, and characterization of focal lesions (10).

Second, dual-echo 3D GRE is used for in-phase and out-of-phase images at two echo times where fat and water signals add or cancel. This sequence is typically acquired in a single breath-hold and is widely used for tissue characterization, particularly for identifying microscopic fat in the liver (11, 12).

Third, multi-echo 3D GRE is used for chemical-shift-encoded fat/water separation. This sequence typically acquires six echoes to enable robust estimation of proton-density fat fraction (PDFF) and R2*, which are important for assessment of hepatic steatosis and iron overload, respectively (13–17). To keep the acquisition within a single breath-hold, multi-echo 3D GRE is often performed at lower spatial resolution than single-echo 3D GRE because of its longer TR. Despite this, it is clinically valuable for assessing diffuse liver disease such as fatty liver disease, where high spatial resolution is often less critical. Recent studies suggest that switching from 3D multi-echo GRE to 2D multi-echo GRE may provide additional advantages for quantitative chemical-shift-encoded imaging (18, 19), which will be discussed in Section 5.

### 3.3 Diffusion-Weighted Imaging (DWI)

DWI is an essential component of abdominal imaging (20–22). Diffusion encoding applies bipolar gradients to sensitize the signal to microscopic water motion, which suppresses signal from tissues with relatively unrestricted diffusion while preserving signal in tissues with restricted diffusion (e.g., tumors or inflammation). Because diffusion encoding increases sequence duration, routine abdominal DWI is usually performed with single-shot EPI (ss-EPI) to maximize efficiency. Similar to SSFSE, ss-EPI acquires each slice within a single long readout (EPI echo train, ~50-150ms), which makes each individual diffusion-weighted image relatively tolerant of intraframe respiratory motion.

In practice, abdominal DWI requires multiple b-values and diffusion directions, and multiple signal averages (NEX) are often needed to ensure adequate signal-to-noise ratio (SNR), especially at high b-values. These requirements make breath-holding difficult, and therefore, abdominal DWI is often performed during free-breathing at many institutions, although breath-hold and respiratory-triggered acquisitions are also considered in some settings (21, 23). With free-breathing DWI, individual diffusion-weighted image may be relatively motion-free because of its efficient data acquisition. However, images across slices, b-values, diffusion directions, and averages may correspond to different respiratory states, which results in slice-to-slice inconsistency and inter-image misalignment. This directly affects downstream processing, including combination of different diffusion directions and averages, as well as voxel-wise estimation of apparent diffusion coefficient (ADC) maps (23–26).

At present, there is no universally accepted guideline defining an optimal motion management strategy for abdominal DWI, and implementation varies across institutions. In this review, discussion of DWI motion sensitivity and motion management primarily reflects the widely used free-breathing, multi-average approach.

### 3.4 Respiratory Motion Sensitivity of Routine Abdominal MRI Sequences

Respiratory motion affects abdominal MRI sequences in different ways depending on their acquisition strategies. A common feature of current default abdominal MRI protocols is that nearly all sequences use Cartesian sampling. Yes, ss-EPI can also be considered Cartesian, although its effective trajectory may deviate from the ideal path and often requires correction. With Cartesian sampling, respiratory motion is typically manifested as ghosting and spatial blurring in 3D imaging and multi-shot 2D imaging, and as slice-to-slice inconsistencies, inter-image misalignment, and motion-averaged blurring in free-breathing single-shot 2D imaging.

Motion sensitivity is substantially different between 2D and 3D acquisitions. All 3D Cartesian sequences are inherently sensitive to respiratory motion because a full 3D volume is acquired over multiple respiratory cycles. Therefore, even modest respiratory motion can introduce blurring and/or ghosting artifacts, with increasing severity for larger respiratory displacement, as shown in **Figure 3**. Respiratory motion can also cause fat/water separation errors in multi-echo 3D GRE (27, 28). These challenges affect both 3D T1-weighted GRE and 3D T2-weighted FSE, making reliable breath-holding or alternative respiratory management strategies essential.

Single-shot 2D sequences, such as SSFSE and ss-EPI, acquire each slice within a short window (e.g., within 1 second), which reduces intraframe motion sensitivity. However, during free-breathing, slices are acquired sequentially over many breathing cycles, each at a different respiratory state. This results in slice-to-slice misalignment and inconsistent organ boundaries across different slices, as shown in **Figure 2** for SSFSE imaging and **Figure 4** for DWI with ss-EPI. In free-breathing DWI, these inconsistencies are further complicated by variation of contrast and SNR across b-values, diffusion directions, and averages, making inter-image misalignment a major limitation even though individual ones may appear motion-free.

In addition to inter-image misregistration across different individual diffusion-weighted images, abdominal DWI is also intrinsically sensitive to respiratory motion during diffusion encoding. Strong diffusion-sensitizing gradients make the signal highly sensitive to phase errors induced by both respiratory and cardiovascular motion, which can lead to signal dropout, ghosting, and quantification bias. Sequence-level approaches such as gradient moment nulling (e.g., M1 and M2 compensation) have been proposed to mitigate this effect (29–32) but require prolonged echo times and additional optimization, which have limited clinical adoption to date. A detailed discussion of diffusion-encoding motion compensation is beyond the scope of this review, which focuses on respiratory motion management at the acquisition and reconstruction levels.

In contrast, multi-shot 2D FSE is sensitive to both intraframe and inter-slice motion. Motion across shots corrupts phase inconsistencies within a slice, while variations between slices introduce additional inconsistency. As a result, multi-shot sequences typically require reliable breath-holding or other effective respiratory compensation strategies. Failed breath-hold can result in severe motion artifacts, as shown in **Figure 5**, which also indicates that multi-shot 2D FSE is more sensitive to respiratory motion compared to 3D GRE imaging because of its longer TR between shots and increased sensitivity to inter-shot inconsistencies.

In summary, the nature of motion sensitivity varies across sequences. For 3D imaging and multi-shot 2D imaging, motion compensation strategies should focus more on mitigating intraframe motion that cause blurring and/or ghosting. For single-shot 2D imaging, particularly free-breathing DWI, major limitations arise from slice-to-slice inconsistency and inter-image misalignment, for which robust image alignment or registration-based methods may be especially beneficial. Among different routine abdominal MRI sequences, free-breathing ss-EPI DWI may arguably be the one that could benefit most from improved respiratory motion management, especially given its relatively inconsistent image quality in clinical practice.

## 4. Classical Methods for Respiratory Motion Management in Abdominal MRI

This section summarizes the classical methods routinely used in clinical abdominal MRI for respiratory motion management, including breath-holding, accelerated acquisition, respiratory triggering, motion-averaged acquisition, and motion-robust acquisition.

### 4.1 Breath-Holding

Breath-holding remains the most widely used and effective approach to avoid respiratory motion artifacts in clinical abdominal MRI. When patients are able to hold their breath reliably, breath-hold imaging is simple, efficient, and robust. However, the requirement for breath-hold is different between 2D and 3D acquisitions.

In 2D imaging, acquisition can typically be divided into multiple breath-holds using the concatenation strategy described earlier. With this approach, only a subset of slices is acquired during each breath-hold, therefore reducing the breath-hold duration required for each acquisition. The trade-off is that the patient needs to perform an increased number of breath-holds to cover the entire imaging volume.

In contrast, all k-space samples contribute to a single reconstructed volume in 3D imaging. This is different from multi-slice 2D acquisitions, where slices are acquired independently. As a result, a single, continuous breath-hold is required to acquire the full 3D dataset, which often necessitates longer breath-holding than 2D imaging. To make such breath-holding feasible, spatial resolution may need to be reduced, particularly for sequences with long TRs such as multi-echo 3D GRE.

### 4.2 Accelerated Acquisition

Accelerated acquisition may seem distinct from motion management, but it is in fact one of the most effective strategies for reducing motion artifacts in clinical practice (33, 34) and is used in almost every abdominal MRI exam. Fast imaging directly shortens the breath-hold duration and makes it easier for patients to better cooperate with the exam. Even in free-breathing scans, faster acquisition helps reduce total scan time and thus decrease the likelihood of respiratory drifts or cycle-to-cycle motion variability. Accelerated imaging can also shorten echo trains in SSFSE and ss-EPI sequences, which in turn reduces T2/T2* blurring or geometric distortion. Early feasibility studies suggest that highly accelerated acquisition may allow fast 3D FSE MRCP within a single prolonged breath-hold to achieve improved scan efficiency (35–37), although such approaches are not widely available in routine practice yet. In addition, increased imaging speed can also be traded for higher spatial resolution without prolonging scan time.

Common acceleration techniques in clinical abdominal MRI include partial Fourier imaging, parallel imaging, compressed sensing, simultaneous multi-slice imaging, and more recently, deep learning-based reconstruction (34). All of these methods play an essential role in motion management by minimizing the acquisition window over which respiratory motion can affect the data.

**4.3 Respiratory Triggering with External or Internal Motion Signals**

When patients cannot perform reliable breath-holds, respiratory triggering provides an alternative motion management strategy (2, 38, 39). Triggering can be implemented using external devices (e.g., respiratory bellows) or internal motion surrogates (e.g., image-based navigators). During normal breathing, motion is monitored continuously and data are acquired only when the motion signal satisfies a predefined condition, commonly near end-expiration. This increases consistency of respiratory phase in the acquired data, so motion artifacts can be effectively mitigated.

Despite its benefits, respiratory triggering has important limitations. The major drawback is reduced scan efficiency and prolonged scan times, since acquisition occurs during only a fraction of the respiratory cycle and the scanner remains idle for the rest. Scan duration may also become unpredictable due to variability in breathing patterns. In addition, triggering is not compatible with 3D GRE sequences that require continuous acquisition to maintain steady-state magnetization, and it is also not suitable for DCE-MRI, where rapid imaging is required to capture contrast dynamics. In current clinical practice, respiratory triggering is most commonly used in 3D MRCP (40, 41), where its long scan time makes breath-holding difficult. Triggering has also been shown to improve image quality in abdominal DWI (25, 26), but this is less preferred in clinical practice due to longer scan time (21).

### 4.4 Free-Breathing Motion Averaged Acquisition

Motion-averaged acquisition, which repeats data acquisition for multiple times and then averages the resulting images, is another simple but clinically useful motion mitigation strategy when breath-holding is not feasible (3, 42). Because respiratory motion varies across repetitions, averaging effectively reduces structured motion artifacts and translate them into spatial blurring, which may be more clinically acceptable than pronounced ghosting. **Figure 6** compares breath-hold acquisitions (left column) with single-average and five-average free-breathing acquisitions (middle and right columns) for both 3D GRE and multi-slice 2D FSE. Motion averaging with increased repetitions reduces motion artifacts present in single-average free-breathing images but increases blurring and scan time.

In general, motion-averaged acquisition is more commonly used for multi-shot 2D FSE than 3D GRE. This is because 3D GRE is often used for DCE-MRI where rapid acquisitions are required to capture contrast enhancement. As described above, motion-averaged acquisition is routinely used for DWI. However, the main benefit of averaging in DWI is for SNR improvement rather than for motion-artifact suppression, since ss-EPI is relatively tolerant of intraframe motion. In fact, averaging diffusion-weighted images acquired at different respiratory positions results in additional blurring, which is an inevitable trade-off for SNR purpose, as presented in **Figure 4**.

### 4.5 Motion-Robust Acquisition

Non-Cartesian sampling is known to be more tolerant of motion than Cartesian sampling, in part because repeated sampling of k-space center introduces a motion-averaging effect and spreads motion-induced phase errors more broadly (43–46). In clinical abdominal MRI, radial and PROPELLER (Periodically Rotated Overlapping ParallEL Lines with Enhanced Reconstruction) are the two most commonly used motion-robust alternatives. Radial sampling, usually implemented using the stack-of-stars trajectory for 3D GRE (43, 44), has been extensively demonstrated as a motion-robust option for patients who cannot hold breath-holds in abdominal MRI, as shown in **Figure 7a**. PROPELLER FSE has long been a motion-robust alternative to multi-shot 2D Cartesian FSE. Like radial sampling, the rotating blade acquisition in PROPELLER inherently reduces in-plane motion artifacts (46), as shown in **Figure 7b**.

Despite their advantages and increasing clinical adoption, both radial sampling and PROPELLER imaging have important limitations. First, they are motion-robust rather than motion-free. As shown in **Figure 7**, both radial sampling and PROPELLER sampling are prone to motion-induced image blurring. Second, standard radial and PROPELLER sampling are less efficient than Cartesian acquisition and often require more data to satisfy the Nyquist criterion. As a result, the scan time of these two sequences is typically longer unless advanced acceleration methods are used. Third, as shown in **Figure 8**, trajectory-specific artifacts, such as streaking in radial imaging, remain a significant concern for routine use (44), and these artifacts make it less robust when applied in non-axial orientations.

**4.6 Choice of Breath-Hold verse Free-Breathing in Clinical Abdominal MRI**

In current clinical practice, the choice between breath-holding and free-breathing acquisitions depends primarily on patient condition, sequence requirements, and institutional preference. In general, 3D FSE MRCP is typically performed during free-breathing with respiratory triggering. While this prolongs scan time, MRCP is not performed in every abdominal MRI exam and is only needed for patients who require dedicated biliary evaluation. Abdominal DWI is often performed during free-breathing, although breath-hold and respiratory-triggered acquisitions are also used at some institutions. 2D SSFSE may be performed during free-breathing or over multiple breath-holds depending on diagnostic needs, but breath-hold acquisitions are generally more preferred due to consistent slice positions. Standard 3D Cartesian GRE is usually performed in a single breath-hold, unless a motion-robust alternative (such as stack-of-stars GRE) is used. For multi-shot 2D FSE, increasing the number of concatenations is a practical strategy to shorten individual breath-hold at the cost of additional breath-holds and longer total scan time. PROPELLER-based T2 FSE is usually reserved for patients who are unable to perform reliable breath-holds.

In routine practice, breath-hold remains the first-line choice whenever feasible – except for DWI and 3D FSE – because it is fast, efficient, and highly effective in minimizing respiratory artifacts. Free-breathing alternatives are generally used when breath-holding is not possible and often require longer scan time and/or more complex reconstruction. One potential exception is DCE-MRI, which may represent an early application where the practical implementation could shift towards free-breathing acquisition without a scan-time penalty, since the contrast-enhancement window spans several minutes regardless the acquisition choice. Therefore, free-breathing DCE-MRI, particularly when combined with advanced dynamic reconstruction that leverages temporal correlations, may become increasingly attractive for clinical translation, which will be discussed further in Section 5.

## 5. Advanced Respiratory Motion Management Methods for Abdominal MRI: Opportunities and Translational Gaps

This section describes both long-standing and more recent motion management strategies that have potential clinical value but have not been widely translated into routine practice for abdominal MRI. These include registration-based motion correction, motion-resolved and motion-weighted reconstruction, sub-second imaging using ultra-fast acquisition, and deep learning-based motion compensation. For each approach, we discuss opportunities, its clinical relevance, and translational gaps that currently limit clinical adoption.

### 5.1 Registration-Based Motion Correction and Alignment
*Opportunity*

Registration-based motion correction using rigid or non-rigid models is one of the most extensively studied strategies (47, 48). Readers may therefore consider registration a classic, well-established solution for mitigating motion artifacts. Indeed, registration has proven effective in neuroimaging, where head motion is predominantly rigid (49). However, its application to abdominal MRI is considerably more challenging due to the complex, deformable, and irregular respiratory patterns as described in Section 2, which limit the robustness and generalizability of simple rigid motion models (50–52).

Despite these challenges, abdominal MRI is expected to greatly benefit from reliable and accurate non-rigid image registration, particularly for applications affected by slice-to-slice inconsistency or inter-image misalignment, such as SSFSE and free-breathing DWI. Several pilot studies have demonstrated that registration-based correction can reduce inconsistencies and improve the quality of free-breathing liver DWI (23, 53), but these approaches have not been translated into routine clinical practice yet.

*Translational Gap*

Overall, progress in registration-based motion correction for abdominal MRI has been slow, primarily due to substantial variability in patient breathing patterns and the inherent error-prone nature of deformable registration. For DWI, two additional barriers have to be considered. First, inter-image misalignment occurs not only across slices or repetitions but also between images acquired with different b-values and diffusion directions, which have different contrast and SNR. These variations violate the assumptions behind conventional intensity-based registration and can make deformation estimation more difficult. Second, the inherently low SNR of DWI further complicates robust image registration.

In addition, the performance of registration-based motion correction also varies across different organs. The kidneys, for example, often exhibit more rigid-like motion due to their retroperitoneal location and surrounding fascial support (54). As a result, respiratory-induced kidney motion occurs with relatively limited deformation compared to other organs such as the liver and pancreas, and the sharp anatomical boundaries of the kidneys further facilitate image registration. Several studies have demonstrated encouraging results using non-rigid registration for motion correction in kidney MRI (55–57). However, kidney-focused abdominal MRI exams are generally less common in current clinical practice than liver or pancreatic MRI, where respiratory motion is more complex. Overall, while registration-based motion correction holds great promise and is highly desirable in abdominal MRI, substantial technical advances are still needed before it can be reliably used in clinical practice.

## 5.2 From Gating to Motion-Resolved and Motion-Weighted Reconstruction

### *5.2.1 Gating-Based Motion Compensation*

<u>Opportunity</u>

Respiratory gating is a widely used technique in cardiovascular MRI and has also been explored for abdominal MRI. In fact, respiratory-triggered acquisition described earlier can be treated as a form of gated imaging, in which data are acquired only when respiratory motion falls within a predefined window. This approach is commonly referred to as prospective triggering or gating.

Respiratory gating can also be performed retrospectively, in which data are first acquired continuously during free-breathing, and a respiratory motion signal, either obtained from an external device or extracted directly from the acquired data (referred to as self-gating), is used to guide the selection of desired k-space samples for motion-corrected images (58). The main advantage of gating-based strategies is that they do not require strong assumptions about patient-specific motion patterns as in image registration.

A key requirement for retrospective gating is that each k-space segment should carry similar amounts of image information, including both low- and high-frequency samples, such that k-space can still be sufficiently covered after motion-guided data sorting. For this reason, retrospective gating is typically used with golden-angle rotated sampling or its variants implemented using non-Cartesian trajectories or specifically designed 3D Cartesian sampling patterns, such as interleaved acquisitions that repeatedly sample the k-space center while allowing flexible retrospective sorting (44, 59–61).

<u>Translational Gap</u>

Similar to prospective triggering, the main limitation of retrospective gating-based motion compensation is reduced scan efficiency, which often leads to prolonged acquisition time. When respiratory patterns become irregular or unpredictable, the scan time can increase significantly. In contrast to prospective triggering/gating that can be broadly applied across different sequences, retrospective gating typically requires specially designed sampling trajectories. Its main advantage, however, is preservation of the steady state, so that it can be used in sequences like 3D GRE, while prospective triggering disrupts steady-state signal evolution. Overall, retrospective gating does not consistently provide a clear practical advantage over prospective triggering, although self-gating remains an active area of research.

### *5.2.2 Motion-Resolved Image Reconstruction*

*Opportunity*

To overcome the limitations of gating-based approaches, particularly the reduced scan efficiency associated, various methods have been proposed to fully utilize all acquired k-space data with 100% imaging efficiency. An early strategy involved sorting data into different motion phases to generate motion-resolved images, as shown in **Figure 9a-c**. This was followed by image registration to combine these images into a single motion-corrected result (62–69). Although this approach attracted considerable interest, it faces two major challenges. First, it relies on accurate image registration, which is inherently difficult in abdominal MRI. Second, motion-resolve images generated after sorting are often highly undersampled with artifacts (**Figure 9c**), which further complicate image registration. Despite these challenges, motion-guided data sorting enabled explicit sorting of k-space data into different respiratory motion states, and this idea later motivated the development of motion-resolved image reconstruction methods that recover image quality from motion-sorted undersampled images (70–73).

Motion-resolved reconstruction does not require explicit assumptions about the underlying motion model. Instead, it exploits the pseudo-periodic nature of respiration to sort k-space data acquired over multiple breathing cycles into a set of distinct respiratory motion states (**Figure 9a-c**). Images reconstructed from each state exhibit reduced intraframe motion and also form an additional respiratory dimension. While each motion state is undersampled with artifacts, the respiratory dimension generated with data sorting provides new temporal correlations that can be leveraged using advanced reconstruction techniques, such as compressed sensing, to remove undersampling artifacts. This is the idea behind XD-GRASP (eXtra-Dimensional Golden-angle RAdial Sparse Parallel imaging) MRI, which first demonstrated motion-resolved reconstruction for free-breathing dynamic MRI using golden-angle radial sampling and sparse reconstruction (70), as shown in **Figure 9c-d**.

Practical implementation of motion-resolved reconstruction requires several components. First, a respiratory surrogate is needed to guide data sorting (**Figure 9a**). This signal can be derived either from the acquired data, such as through self-navigation from repeatedly sampled k-space center or from an external device. Second, similar to gating-based approaches, each motion state should retain sufficiently uniform k-space coverage after sorting (**Figure 9b**), which is typically achieved using golden-angle rotated sampling schemes that allow flexible retrospective sorting (44). In abdominal MRI, motion-resolved reconstruction has primarily been demonstrated for free-breathing 3D GRE imaging (74) and is also expected to benefit 3D FSE MRCP as a more efficient alternative to respiratory triggering (75).

*Translational Gap*

Motion-resolved image reconstruction faces several practical limitations. First, explicit data sorting into multiple respiratory states is particularly challenging in time-sensitive applications such as DCE-MRI, where image intensity changes rapidly due to contrast enhancement and each phase contains limited data for reliable sorting. Second, accurate and robust respiratory motion signals are required to guide data sorting, which are not always easy to obtain in abdominal MRI, and motion detection becomes more challenging when tissue contrast changes dynamically, such as in DCE-MRI. Third, reconstruction with an additional respiratory dimension increases computational burden and reconstruction time, representing another barrier to routine clinical deployment. Fourth, it requires specially designed acquisition trajectories and is not naturally compatible with routine clinical sequences. As a result, although motion-resolved image reconstruction provides a novel way for handling respiratory motion, its widespread clinical translation has so far been limited.

### *5.2.3 Motion-Weighted Image Reconstruction*
*Opportunity*

Motion-weighted reconstruction, also known as soft-gating, mitigates motion artifacts without explicit data sorting (76–78). Instead of resolving data into distinct motion states, soft-gating assigns different weights to k-space samples according to their respiratory phase. For example, data acquired near end-expiration can be weighted more heavily with more contribution to the final image, while data acquired during inspiration are weighted less. This implicit motion-compensation strategy has been shown to effectively reduce motion artifacts without generating an extra motion dimension or substantially increasing reconstruction time (76–81), as shown in **Figure 10**.

*Translational Gap*

Despite these advantages, motion-weighted reconstruction shares several major limitations with motion-resolved reconstruction. Its performance strongly depends on the availability of accurate and robust respiratory motion signals, and it also requires special sampling trajectories. These factors make the approach sensitive to motion detection errors and challenging to implement in routine clinical sequences.

**5.3 Sensor-Guided Free-Breathing Imaging**

*Opportunity*

Many free-breathing approaches rely on accurate estimation of respiratory motion, which can be particularly challenging when underlying image intensity changes over time, such as in DCE-MRI. In recent years, non-contact motion sensors have emerged as an active area of development, and several vendor-integrated solutions are now available, such as PilotTone or Beat Sensor (Siemens Healthineers) (82, 83) and VitalEye (Philips Healthcare) (84). These technologies provide respiratory traces without requiring specific patient cooperation and can be used for motion gating or data sorting. A key advantage of these motion sensors is that they are directly integrated into MRI systems, which enables relatively seamless incorporation into clinical workflows.

*Translational Gap*

Despite growing interest, sensor-guided motion monitoring does not directly enable motion compensation and must be combined with dedicated correction algorithms.

As such, many limitations associated with existing motion compensation methods remain. In addition, most existing sensors only provide a single global respiratory signal at each time point, which may be insufficient to capture irregular breathing patterns, spatially heterogeneous motion, or baseline drift. Future sensor advances that provide richer spatial motion information may improve the robustness and accuracy of respiratory detection and facilitate more effective prospective or retrospective motion correction.

### 5.4 Sub-Second Imaging Using Ultra-Fast Acquisition

*Opportunity*

One fundamental reason MRI is sensitive to respiratory motion artifacts is its relatively slow imaging speed. If images can be acquired sufficiently fast (e.g., with a temporal footprint of less than one second), the impact of respiratory motion can be reduced because breathing occurs on a much longer timescale. Single-shot sequences are inherently more robust to intraframe motion due to this reason, as discussed in Section 3. While this principle is straightforward for 2D imaging, extending it to 3D acquisitions has been historically more challenging with conventional reconstruction methods.

Recent advances in image reconstruction have demonstrated that a full 3D volume can be acquired in less than one second for DCE-MRI by aggressively accelerating each contrast phase and exploiting temporal correlations along the contrast-enhancement dimension (85–88). One example of such techniques is GRASP-Pro (Golden-angle RAdial Sparse Parallel imaging with imProved performance), which integrates multicoil compressed sensing with a low-rank subspace model to enable continuous free-breathing DCE-MRI with a sub-second temporal footprint (86), as shown in **Figure 11**. Acquiring each 3D volume within one second reduces intraframe motion artifacts and also eliminates the need for explicit motion detection and motion compensation. Moreover, this imaging strategy can be more robust to irregular breathing or bulk movement compared to motion-resolved reconstruction, as shown in **Figure 12** (89).

While this idea is relatively new in abdominal MRI, similar concepts have long been explored in cardiac MRI, where real-time ungated imaging has proved valuable for patients with arrhythmias by rapidly acquiring each cardiac phase (90–93). Related ideas have also been applied in other applications. For example, a recently proposed quantitative fat/water imaging approach based on 2D chemical-shift-encoded acquisitions with flip-angle modulation (FAM) acquires each slice within a short temporal window (typically <1 s) and repeats the acquisition with different flip angles (94, 95). Compared to conventional multi-echo 3D GRE, which is more sensitive to respiratory motion due to longer acquisition window, 2D FAM enables robust free-breathing imaging with minimal intraframe motion while also correcting for residual T1 bias (94).

*Translational Gap*

Sub-second volumetric imaging is particularly attractive for free-breathing DCE-MRI, where strong temporal correlations can be leveraged to aggressively accelerate each contrast phase. However, several practical challenges remain. First, these approaches often impose substantial computational demands, particularly when advanced iterative or model-based reconstructions are needed. Second, the large number of reconstructed volumes (e.g., one 3D volume per second or faster) necessitates automated identification of clinically relevant contrast phases. Third, the effective use of temporal resolution in sub-second dynamic imaging requires specially designed trajectories (e.g., golden-angle rotated sampling), similar to motion-resolved reconstruction.

In addition, while sub-second imaging effectively mitigates intraframe motion, it does not eliminate inter-frame respiratory variation. This may affect quantitative analysis where temporal motion consistency is required, although it is less likely to limit current clinical practice, where image interpretation remains largely qualitative. Integration of deep learning into both reconstruction and image processing may provide promising solutions to address these barriers.

**5.5 Deep Learning-Based Motion Compensation**

*Opportunity*

Deep learning has been increasingly adopted for various medical imaging applications, including clinical abdominal MRI, where it has been successfully deployed to accelerate data acquisition and improve image quality (96). However, the use of deep learning specifically for motion compensation in abdominal MRI remains at a relatively early stage. In this section, we discuss three major categories of deep learning techniques relevant to motion management: (a) deep learning-based reconstruction for accelerated acquisition, (b) deep learning-based motion artifact suppression, and (c) joint deep learning reconstruction and motion compensation.

Deep learning-based reconstruction is now widely implemented clinically (97). By enabling higher acceleration with reduced scan time, these methods shorten the acquisition window during which respiratory motion can corrupt data. Although not a direct motion correction strategy, scan time reduction is one of the most effective ways to mitigate motion artifacts.

Beyond acceleration, deep learning-based motion artifact suppression has been explored as a post-processing strategy to reduce blurring and ghosting caused by respiratory motion. In this paradigm, neural networks are trained to map motion-corrupted images to artifact-reduced outputs, and pilot studies have shown promising results (98–100).

More advanced approaches integrate deep learning with image reconstruction and motion modeling, enabling simultaneous estimation of motion fields/vectors and recovery of motion-corrected images from undersampled data. These methods often integrate physics-based constraints and explicit motion models to improve performance (101–106).

*Translational Gap*

Despite their promise, deep learning-based motion artifact suppression methods are largely data-driven and do not explicitly incorporate the physics of respiratory motion. This raises concerns regarding generalization across different patients and breathing patterns, as well as risks of hallucination or inadvertent suppression of subtle pathology. As a result, these approaches remain largely exploratory at the current stage.

Similarly, although joint reconstruction and motion compensation frameworks are conceptually attractive, practical implementation is often complex. Accurate estimation of non-rigid abdominal motion from highly undersampled data is intrinsically challenging, particularly when image contrast varies rapidly. Most studies are limited to proof-of-concept studies with small datasets and lack large-scale validation to demonstrate robustness and reproducibility.

Overall, deep learning holds substantial potential for improving motion management in abdominal MRI. However, unlike deep learning-based reconstruction that enables accelerated data acquisition, which is now routinely available, deep learning-based motion compensation remains in an early developmental phase. Major translational challenges include robust generalization across heterogeneous respiratory patterns, prevention of hallucination, and reliable modeling of complex abdominal motion. Addressing these challenges will require large-scale validation across diverse patient cohorts, and this will be essential before routine clinical integration is possible.

# 6. Future Directions for Respiratory Motion Management in Abdominal MRI

By this section, readers are expected have a clear understanding of the clinical needs for effective respiratory motion management in abdominal MRI and cause of the gap between various motion compensation techniques and routine clinical translation. This section summarizes sequence-specific considerations and highlights future research directions that may better align technical developments with clinical workflows.

## 6.1 Sequence-Specific Consideration and Clinical Relevance

As described in Section 3, the requirements for respiratory motion management vary substantially across abdominal MRI sequences. For 2D SSFSE and DWI, reliable methods to align different slices and correct inter-image misalignment in free-breathing acquisitions remain major unmet needs. For these two sequences, faster acquisition is also desired, which can further shorten echo trains to reduce T2/T2* blurring and geometric distortion or increase spatial resolution.

For DWI, improved alignment across b-values, diffusion directions, and averages could improve the reliability of abdominal DWI by reducing motion-averaging blurring, which is particularly important for small motion-sensitive structures such as the pancreas. In a recent study, deep learning has been applied to automatically identify relatively consistent diffusion-weighted images and selectively combine them for averaging (107).

This data-driven frame selection approach improves image quality and reduces motion-induced blurring compared to conventional averaging of all repetitions.

For multi-shot 2D and 3D FSE sequences, methods that increase imaging speed and reduce scan time could provide high clinical value. Faster imaging can shorten individual breath-hold duration without requiring additional concatenations and could also make single-breath-hold 3D FSE feasible while maintaining image quality comparable to respiratory-triggered acquisitions. This would substantially improve efficiency for MRCP and other 3D FSE applications. When 3D FSE is performed during free-breathing, approaches that increase scan efficiency, such as motion-resolved or motion-weighted reconstruction combined with deep learning-based reconstruction, may enable more consistent acquisition time compared to respiratory triggering or gating that acquire data during only a fraction of the respiratory cycle.

For 3D GRE sequences, DCE-MRI may be the first application to move from breath-hold to free-breathing acquisition, particularly through sub-second volumetric imaging that has demonstrated strong robustness, once remaining practical challenges, such as reconstruction speed and the automated selection of clinically relevant contrast phases, are addressed. Importantly, moving DCE-MRI from breath-holds to free-breathing does not necessarily impose a scan-time penalty because the acquisition needs to cover the entire contrast enhancement duration regardless of acquisition strategies. In contrast, for other 3D GRE applications such as multi-echo fat/water imaging or hepatobiliary-phase imaging with Gadoxetate disodium (Eovist/Primovist), breath-hold remains preferred due to short scan time. Free-breathing methods are likely to remain reserved for patients unable to perform reliable breath-holds unless future developments reduce free-breathing scan time to match current breath-hold protocols.

We propose two practical recommendations for the development of future motion management techniques. First, simple approaches tend to be more robust. For example, methods that avoid complex motion modeling or reduce dependence on patient-specific motion assumptions typically perform more consistently, especially given the substantial variability of abdominal motion among patients. As a result, the generalization and translation of these methods into routine clinical workflows are expected to be easier. Second, potential free-breathing approaches for first-line clinical use should not substantially prolong total scan time compared to current clinical sequences, as scan efficiency is a key factor affecting selection of different imaging strategies, especially at the departmental and institutional levels. If this requirement cannot be met, free-breathing techniques will likely remain primarily reserved for patients unable to perform reliable breath-holds.

### 6.2 Intelligent Selection of Breath-Hold or Free-Breathing Acquisitions

As discussed earlier, breath-hold acquisitions are typically the first choice for most sequences in abdominal MRI, and free-breathing alternatives are used only when a patient cannot hold breath. In many exams, however, this decision is often made after several failed breath-hold attempts, which results in unnecessary delays and reduced workflow efficiency. Intelligent scan-selection based on artificial intelligence could guide technologists in choosing the appropriate strategy, either breath-hold or free-breathing, at the start of the examination. Such intelligent triage can incorporate patient factors, prior imaging from the same subject, or even features extracted from localizers, and it holds great potential to reduce repeated acquisitions and improve efficiency and consistency.

### 6.3 Advanced Image Denoising

Although denoising may seem distinct from motion management, it can indirectly reduce motion artifacts by enabling shorter scans and reducing the need for signal averaging. As discussed in Section 3, DWI typically requires multiple signal averages to ensure adequate SNR. Because these averages are often acquired under free-breathing, the resulting averaged images can suffer from blurring due to the combination of images acquired at different respiratory states (**Figure 4**) As a result, reducing the number of averages could decrease such blurring but at the cost of lower SNR.

Advanced denoising, particularly self-supervised deep learning approaches, offers a promising way to restore SNR while allowing fewer averages (108, 109). Such approaches could improve the quality of DWI (110), especially for small or motion-sensitive structures, and may extend to other sequences that require multiple averages. Furthermore, joint image reconstruction and denoising may provide additional benefit to improve image quality and scan efficiency together (111).

## 6.4 Improved Radial and PROPELLER Imaging

Radial and PROPELLER acquisitions have both demonstrated great clinical value in abdominal MRI by enabling free-breathing imaging. However, these sequences are typically used as backup options rather than default choices to date due to several factors including reduced imaging efficiency, longer scan time, challenges in maintaining consistent image quality, and trajectory-specific artifacts such as streaking in radial imaging (**Figure 8**).

Advances in reconstruction methods that improve image quality and robustness could allow radial and PROPELLER imaging to be more time-efficient while providing improved diagnostic quality in all patients regardless of their breath-hold capacity. Such progress could shift these approaches from secondary options to preferred acquisition strategies. Radial sampling, in particular, holds strong potential for free-breathing DCE-MRI, as demonstrated by motion-resolved reconstructions using XD-GRASP (70) and sub-second imaging techniques using GRASP-Pro (86). With further improvements in reconstruction speed, artifact suppression, and robustness, radial imaging may achieve broader clinical adoption in future abdominal protocols.

## 6.5 Further Advances in Sub-Second Imaging Approaches

As discussed in Section 5, sub-second imaging approaches hold great potential for addressing intraframe respiratory motion in a more robust and generalizable way. A key advantage of this type of approach is that they do not rely on strong assumptions or complex motion models and instead exploit the simple principle that motion artifacts and blurring can be effectively minimized when the data acquisition window is shorter than the timescale of respiratory motion. Existing studies have demonstrated the promise of this imaging strategy in free-breathing DCE-MRI (86), where rapid imaging has led to improved robustness against variable breathing patterns.

This concept can potentially be extended to other sequences as well. For example, in multi-shot 2D FSE acquisitions, one could perform shot-resolved imaging using the idea of motion-resolved image reconstruction, in which images from individual shots are reconstructed separately instead of being combined into a single image. By leveraging temporal correlations across shots, this approach could reduce sensitivity to inter-shot motion and enable more robust free-breathing multi-shot 2D FSE imaging.

**6.7 More Advanced Registration for Motion Correction**

Registration-based motion correction has been discussed throughout this review and remains an important unmet need in abdominal MRI. Despite extensive research, existing registration algorithms are generally not ready for immediate clinical use due to limitations in robustness and accuracy. Emerging deep learning-based registration methods may provide new opportunities for improved speed, robustness, and consistency. They may also better incorporate multi-contrast information, which is particularly relevant for DWI where contrast varies across b-values and directions. Advanced generative approaches, such as diffusion models and other foundation-models, may further improve the performance of image registration. It is also important to realize that registration may be most clinically relevant for correcting inter-image and slice-to-slice inconsistency in abdominal MRI, while intraframe motion in 3D acquisitions may be better addressed through other approaches discussed above, unless registration assumptions are carefully justified.

# 7. Conclusion and Lessons Learned from the Gap in Respiratory Motion Management

Respiratory motion remains a major challenge in abdominal MRI, and the development of effective motion management strategies continues to be an active area of research. This review analyzed and discussed the barriers underlying the gap between technical innovation and clinical translation in respiratory motion management, clarifies the practical clinical needs for addressing the motion challenge in abdominal MRI, and highlights future directions that may better align research advances with clinical workflows and diagnostic priorities.

The translational gaps discussed in the previous sections provide lessons extending beyond respiratory motion compensation in abdominal MRI. Overall, these observations indicate that successful translation of new MRI technologies depends not only on methodological innovation but, perhaps more importantly, on close alignment with practical clinical needs and workflow. The following considerations may help guide future technical development of general MRI technologies.

First, technical innovation should begin with a clear understanding of the clinical problem, the underlying clinical needs, and the limitations of current methods. Active engagement with experienced radiologists and, when appropriate, MRI vendors, can provide valuable insights into unmet clinical needs, practical workflow constraints, and realistic development pathways. In this context, commonly used image quality metrics, such as root mean square error (RMSE), structural similarity index (SSIM), and other related measures, are insufficient to ensure meaningful clinical value, and evaluation of new methods should involve experienced end users (e.g., radiologists) and focus more on diagnostic confidence, robustness in routine workflows, and potential clinical impact when clinical adoption is the goal.

Second, many novel techniques are limited to proof-of-concept studies in controlled settings. Methods aiming for clinical adoption should incorporate a clear pathway towards larger-scale, heterogeneous, and preferably multicenter evaluation to establish robustness and generalizability in clinical environments. Although such studies are inherently challenging and require coordinated efforts among physicists, clinicians, and often institutional stakeholders, they represent an essential step towards ultimate clinical translation.

Third, even technically sophisticated methods may face adoption barriers if they require scanner-specific or institution-specific adjustment to acquisition protocols or reconstruction algorithms. As a result, approaches that are compatible with existing scanners and routine MRI protocols are more likely to demonstrate meaningful clinical value. In this context, open-source dissemination may represent an effective way to facilitate cross-scanner and cross-vendor validation and may therefore accelerate broader translation.

Fourth, robustness to patient variability (e.g., breathing patterns) should be treated as a central consideration in technical development. One lesson from the rapid clinical translation of some prominent MRI techniques, such as parallel imaging and, more recently, deep learning-based reconstruction, is their minimal reliance on patient-specific assumptions. In contrast, many motion management methods depend on strong assumptions about respiratory models, and they are often demonstrated only in small proof-of-concept studies involving healthy volunteers. This limitation can significantly restrict generalizability.

Fifth, successful translation of new MRI techniques into routine clinical practice typically requires close collaboration between academic researchers, clinicians, and industry partners. New models of academic-industry partnership that can facilitate broad clinical evaluation will therefore be crucial to expedite translation of new MRI technologies, including respiratory motion management techniques.

## Acknowledgement

This work was supported in part by the NIH (R01EB030549, R01EB031083, R01DK143170, R21EB032917, and P41EB017183).

# Figures

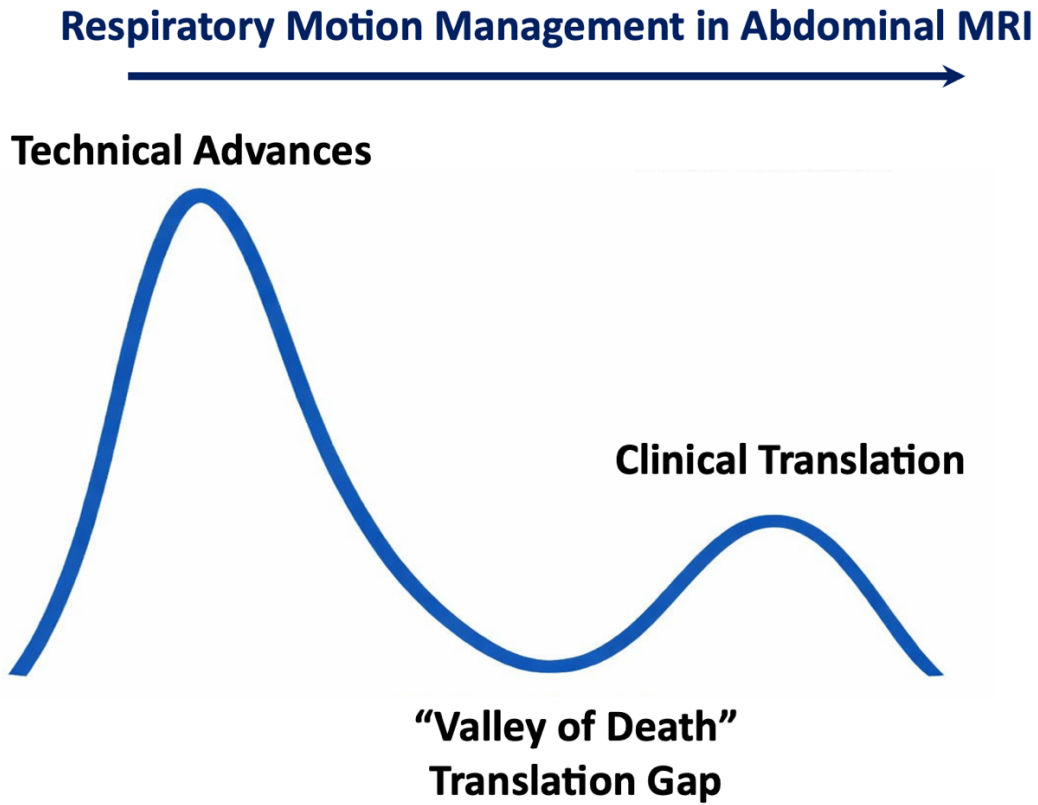

**Figure 1:** Conceptual illustration of the "valley of death" in clinical translation of respiratory motion compensation methods in abdominal MRI. The large peak represents extensive developments of various motion management techniques over past decades, while the much smaller peak reflects the limited number of approaches adopted in routine clinical practice. The valley between the two peaks indicates the significant gap between technical innovation and clinical translation.

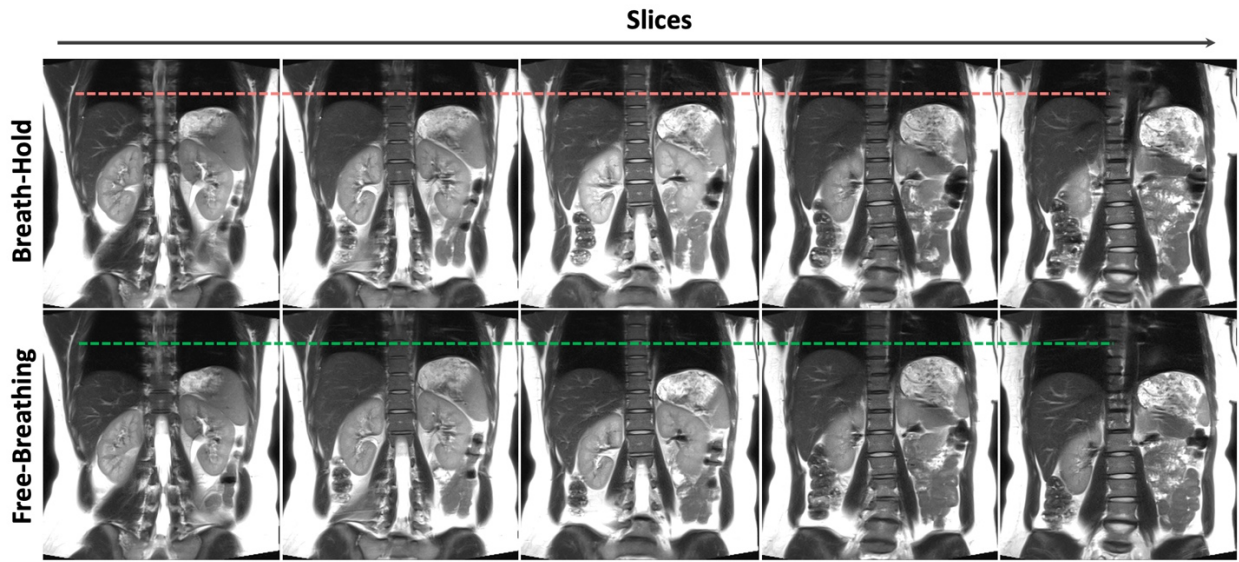

**Figure 2:** Comparison of 2D SSFSE acquisitions during breath-hold and free-breathing. Breath-hold acquisition shows relatively consistent slice positions, while free-breathing acquisition results in respiratory-induced inconsistency across slices as indicated by the green dashed line.

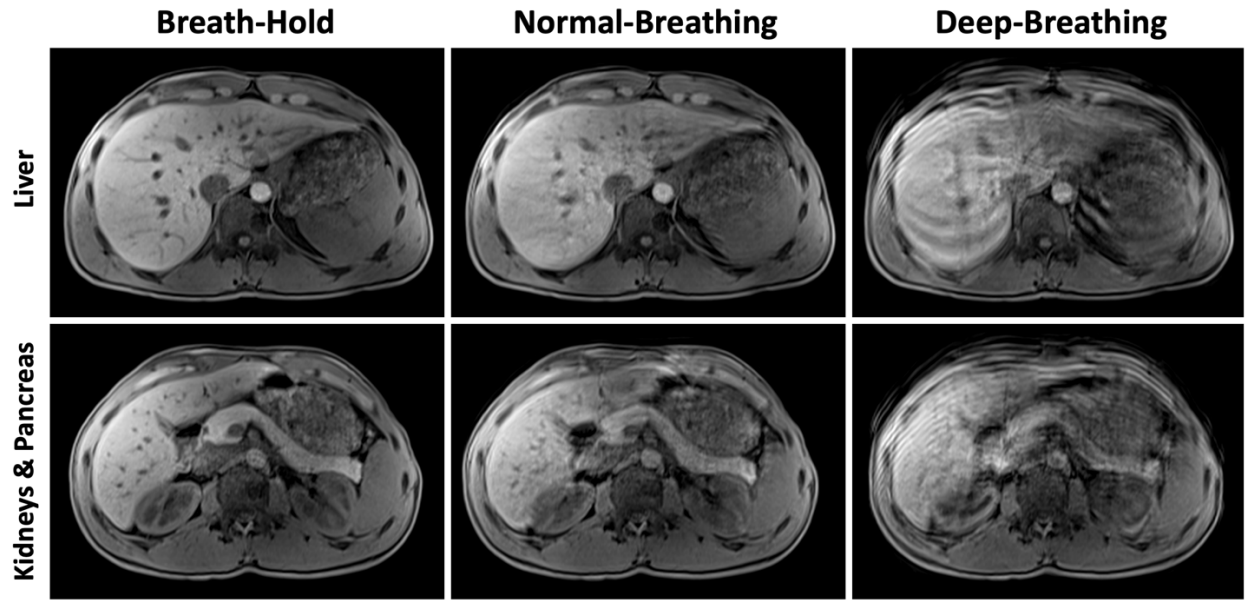

**Figure 3:** Comparison of 3D GRE acquisitions during breath-hold, normal breathing, and deep breathing. Respiratory motion leads to blurring and ghosting along the phase-encoding directions, and larger respiratory displacement produces more severe artifacts.

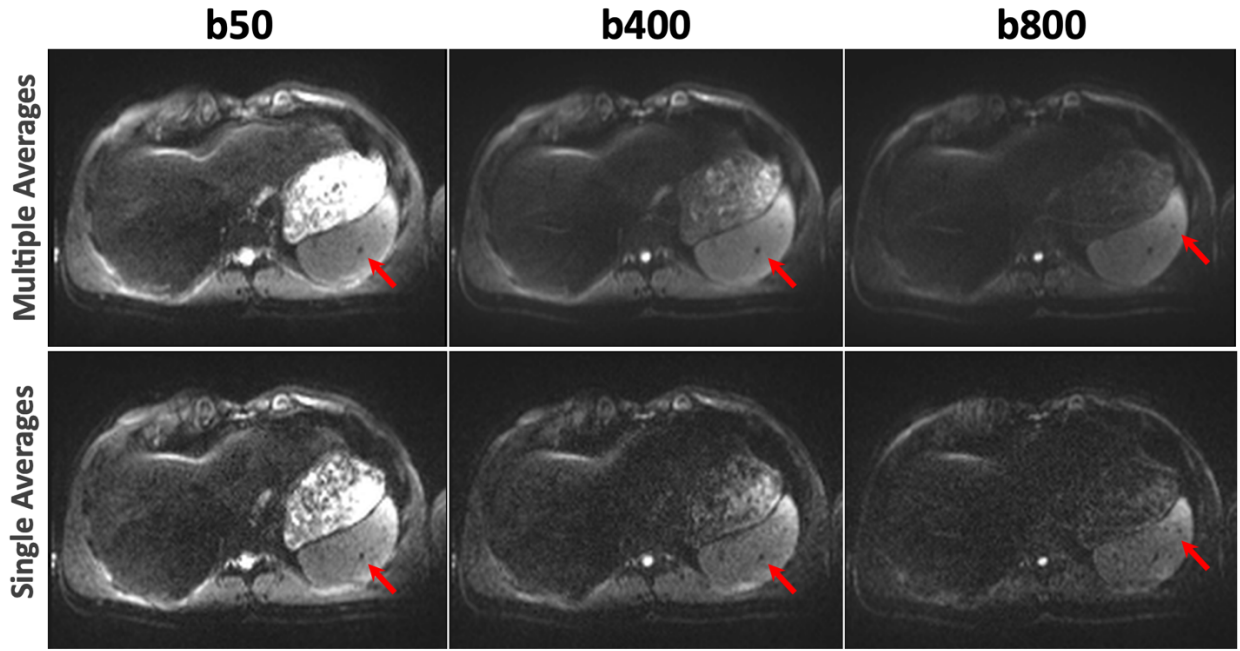

**Figure 4:** Comparison of free-breathing liver DWI with multiple averages (top row) and a single average (bottom row). Although each individual diffusion-weighted image is relatively robust to motion with ss-EPI acquisition, it has low SNR, and multiple signal averages combine images acquired at different respiratory states, thus resulting in spatial blurring, as indicated by the red arrows.

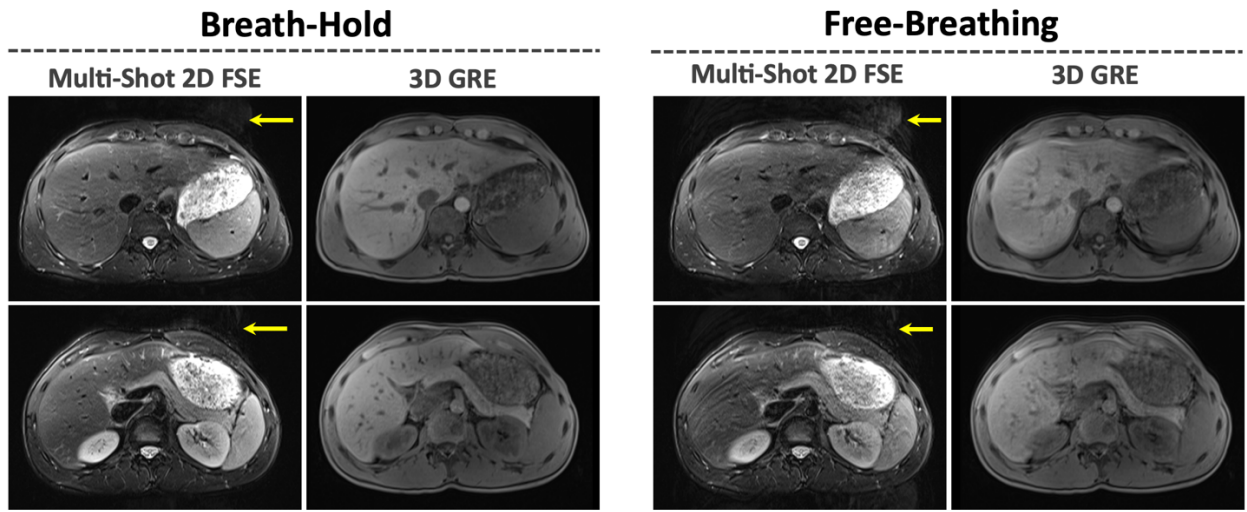

**Figure 5:** Comparison of breath-hold and free-breathing acquisitions in multi-shot 2D FSE and 3D GRE imaging. Free-breathing acquisitions result in blurring and ghosting artifacts that degrade image quality compared to breath-hold acquisitions. Multi-shot 2D FSE exhibits more pronounced motion artifacts than 3D GRE during normal breathing due to longer TR. Yellow arrows indicate that multi-shot 2D FSE is more sensitive to motion than 3D GRE due to the longer TR.

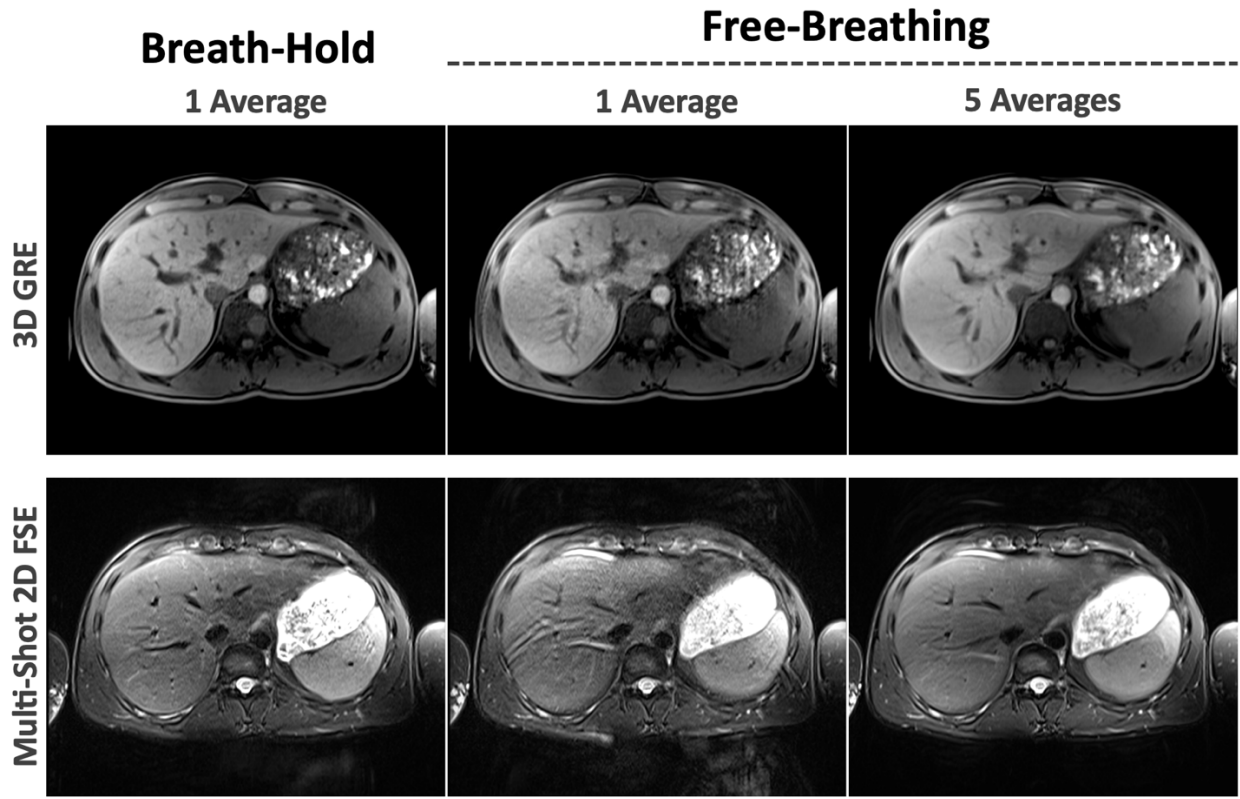

**Figure 6:** Effect of motion averaging during free-breathing for multi-shot 2D FSE and 3D GRE imaging. Increasing the number of averages reduces motion artifacts but introduces spatial blurring and prolongs acquisition time.

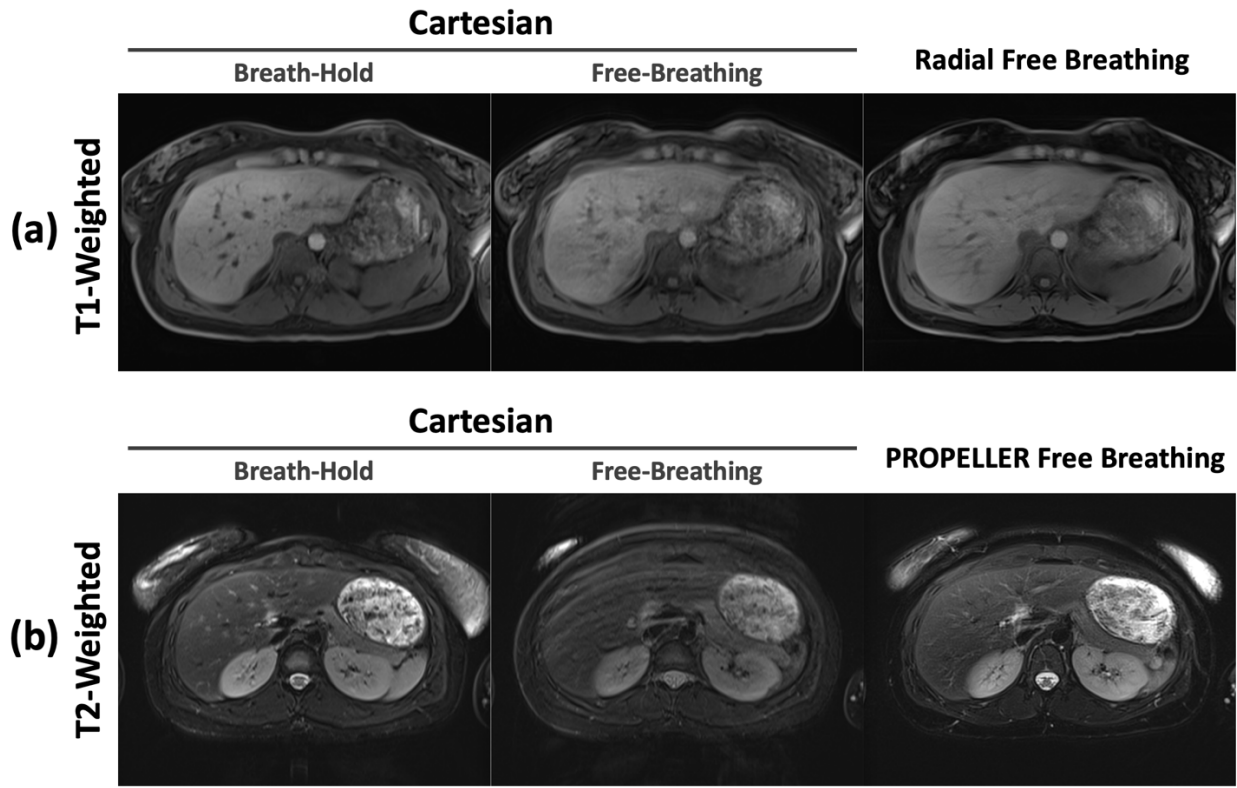

**Figure 7:** Comparison of Cartesian and non-Cartesian (radial and PROPELLER) sampling strategies in free-breathing T1- and T2-weighted liver MRI. Although relatively robust to motion, free-breathing non-Cartesian acquisitions may introduce spatial blurring that degrades image quality.

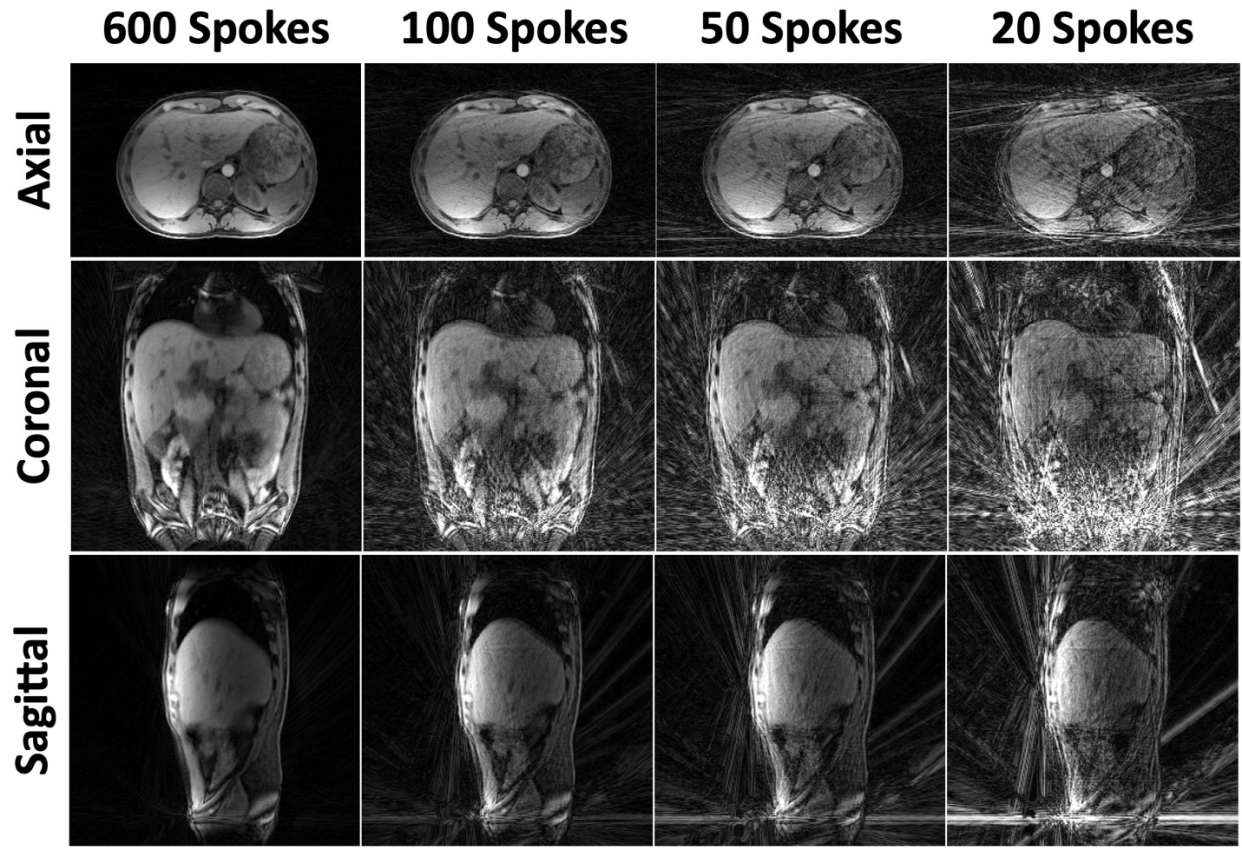

**Figure 8:** Radial sampling (and non-Cartesian sampling in general) is better suited for axial imaging, Coronal and sagittal imaging tend to exhibit more pronounced streaking artifacts with undersampling due to strong high-intensity signal components near the periphery of the field of view. Figure was reproduced from Figure 15 in J Magn Reson Imaging. 2022 Jul;56(1):45-62 with permission from the journal.

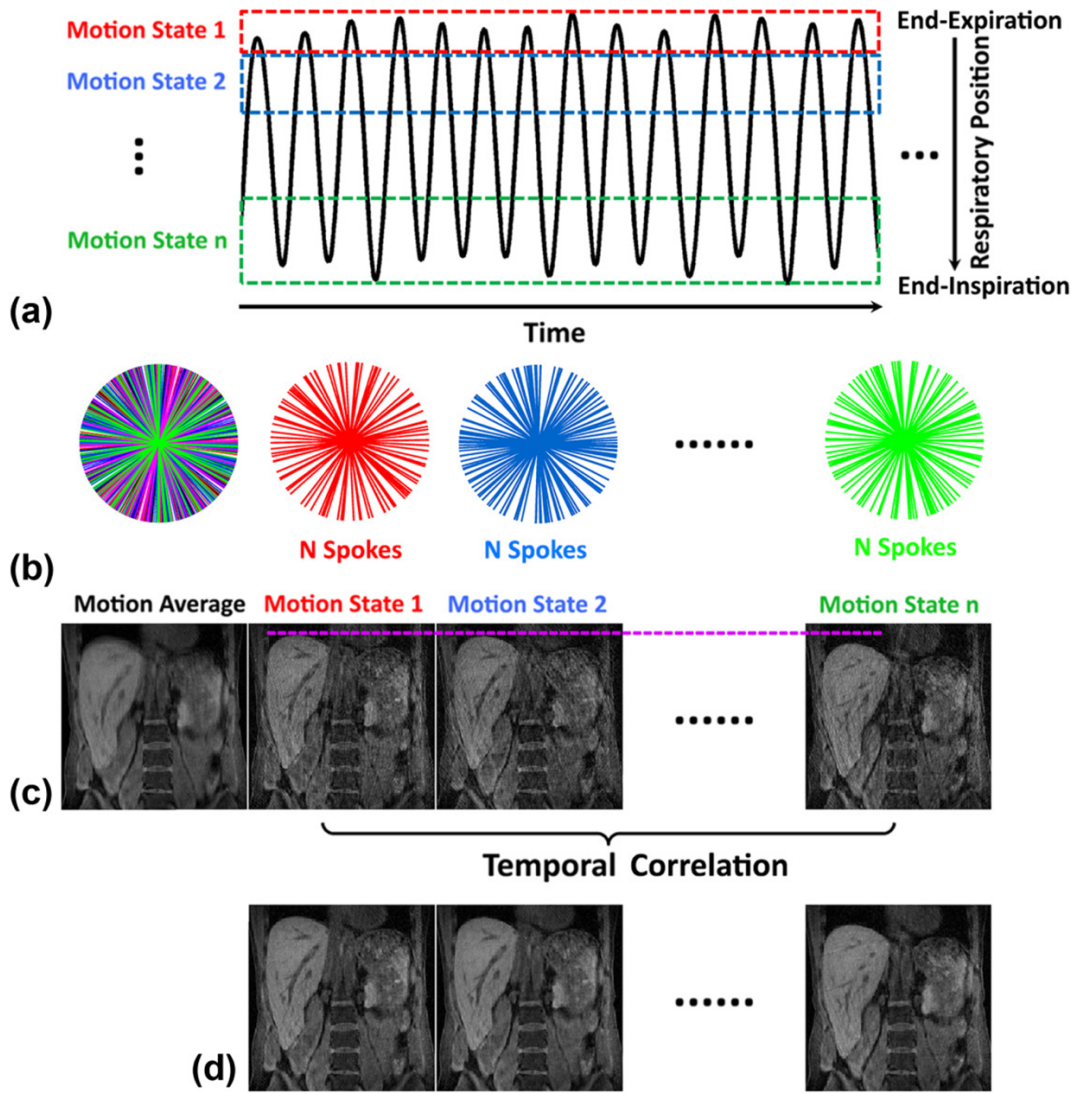

**Figure 9:** (a-b) Motion-guided data sorting using a respiratory motion signal extracted from an external device or from the acquired k-space data (self-navigation). Data can be sorted into multiple motion states spanning from end-expiration to end-inspiration with reduced motion blurring in each sorted image. Golden-angle rotated sampling schemes are typically used to ensure adequate k-space coverage after motion sorting. (c) Despite reduced motion blurring, the sorted images exhibit undersampling artifacts that degrade image quality. (d) Sparsity-based reconstruction can be applied to the sorted data to exploit temporal correlations along the respiratory dimension and generate motion-resolved images with improved image quality. Figure was reproduced from Figure 1 in Magn Reson Med. 2016 Feb;75(2):775-88 with permission from the journal.

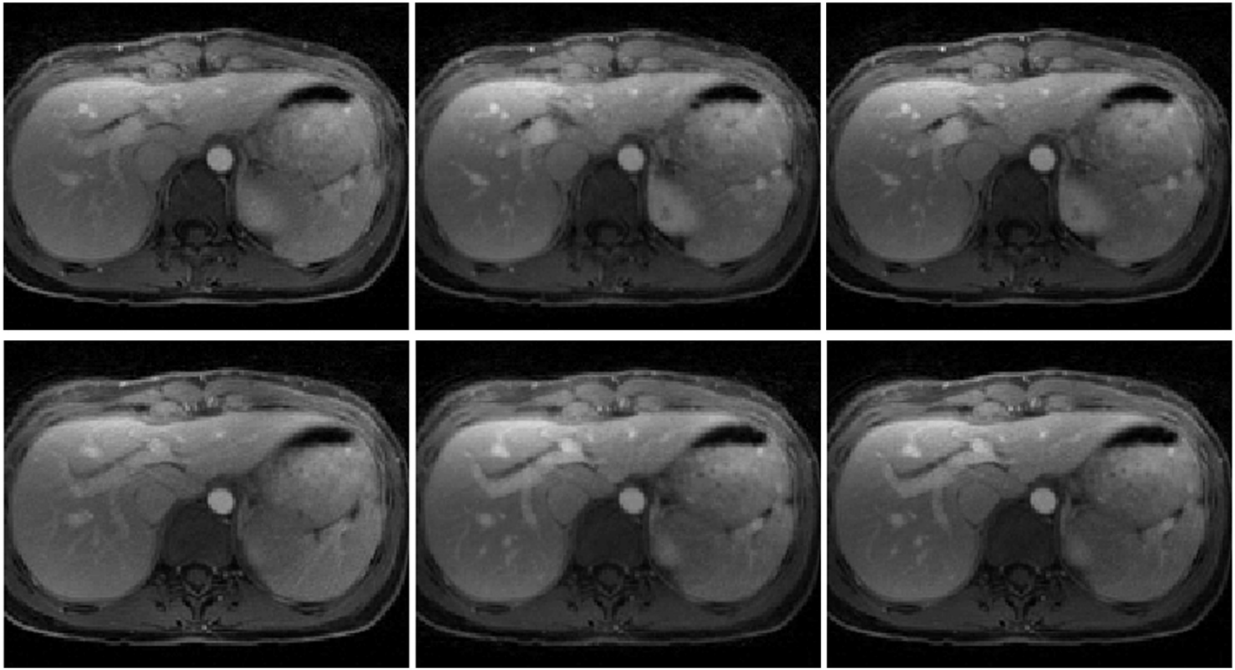

**Figure 10:** Comparison of motion-averaged, motion-resolved, and motion-weighted (soft-gated) reconstructions. Motion-averaged images exhibit noticeable motion blurring, while both motion-resolved and motion-weighted approaches reduce intraframe motion artifacts with comparable image quality.

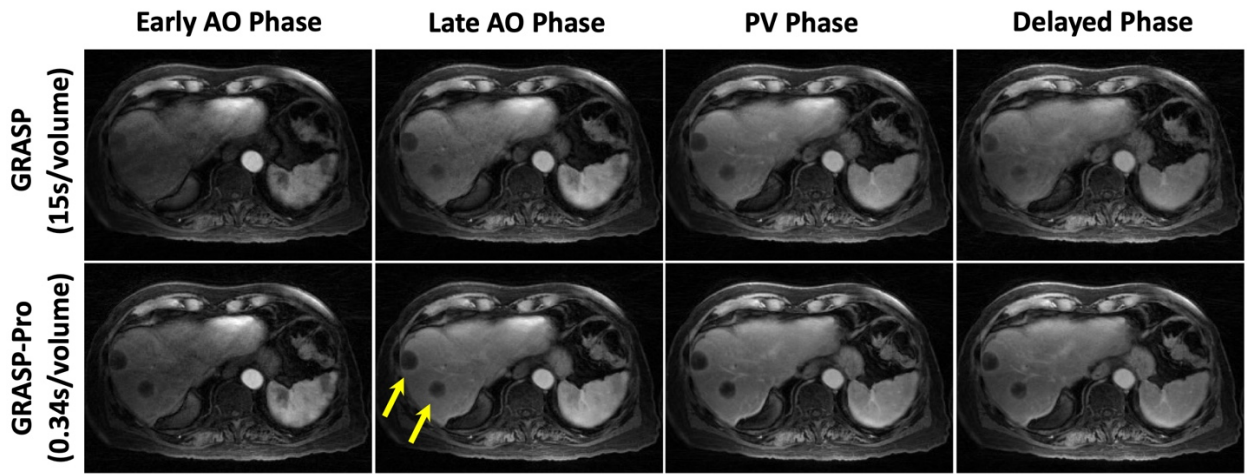

**Figure 11:** Comparison of DCE-MRI reconstructed using GRASP-Pro and standard GRASP. GRASP-Pro enables sub-second temporal resolution that effectively mitigates respiratory motion artifacts and also eliminates the need for explicit motion compensation.

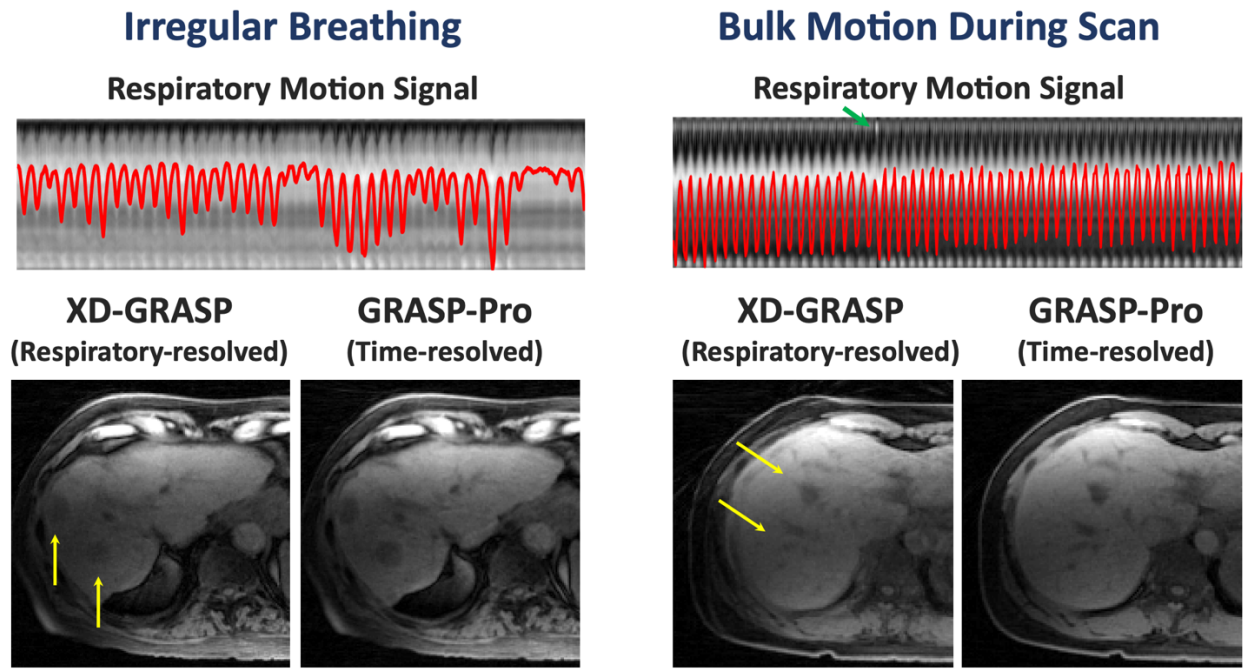

**Figure 12:** Time-resolved, sub-second 4D dynamic MRI using GRASP-Pro demonstrates improved robustness to irregular breathing and bulk motion (indicated by green arrow) compared to motion-resolved reconstruction using XD-GRASP. Yellow arrows indicate motion-induced blurring in the XD-GRASP images. Figure was reproduced from Figure 3 in Magn Reson Med. 2023 Sep;90(3):1053-1068 with permission from the journal.

# References

1. Nepal P, Bagga B, Feng L, Chandarana H: Respiratory Motion Management in Abdominal MRI: Radiology In Training. *https://doi.org/101148/radiol220448* 2022; 306:47–53.
2. Schrauben EM, Lima da Cruz G, Roy CW, Küstner T: Motion Mitigation Techniques for Abdominal and Cardiac MR Imaging. *J Magn Reson Imaging* 2025.
3. Zaitsev M, Maclaren J, Herbst M: Motion Artefacts in MRI: a Complex Problem with Many Partial Solutions. *J Magn Reson Imaging* 2015; 42:887.
4. McClelland JR, Hawkes DJ, Schaeffter T, King AP: Respiratory motion models: A review. *Med Image Anal* 2013; 17:19–42.
5. Foskolou A, Emmanouil A, Boudolos K, Rousanoglou E: Abdominal Breathing Effect on Postural Stability and the Respiratory Muscles' Activation during Body Stances Used in Fitness Modalities. *Biomechanics 2022, Vol 2, Pages 478-493* 2022; 2:478–493.
6. Korin HW, Ehman RL, Riederer SJ, Felmlee JP, Grimm RC: Respiratory kinematics of the upper abdominal organs: a quantitative study. *Magn Reson Med* 1992; 23:172–178.
7. Semelka RC, Kelekis NL, Thomasson D, Brown MA, Laub GA: HASTE MR imaging: description of technique and preliminary results in the abdomen. *J Magn Reson Imaging* 1996; 6:698–699.
8. Sodickson A, Mortele KJ, Barish MA, Zou KH, Thibodeau S, Tempany CMC: Three-dimensional fast-recovery fast spin-echo MRCP: comparison with two-dimensional single-shot fast spin-echo techniques. *Radiology* 2006; 238:549–559.
9. Zhang J, Israel GM, Hecht EM, et al.: Isotropic 3D T2-Weighted MR Cholangiopancreatography with Parallel Imaging: Feasibility Study. *https://www.ajronline.org/* 2012; 187:1564–1570.
10. Do RKG, Rusinek H, Taouli B: Dynamic Contrast-Enhanced MR Imaging of the Liver: Current Status and Future Directions. *Magn Reson Imaging Clin N Am* 2009; 17:339–349.
11. Merkle EM, Nelson RC: Dual gradient-echo in-phase and opposed-phase hepatic MR imaging: a useful tool for evaluating more than fatty infiltration or fatty sparing. *Radiographics* 2006; 26:1409–1418.
12. Fischer MA, Nanz D, Reiner CS, et al.: Diagnostic performance and accuracy of 3-D spoiled gradient-dual-echo MRI with water- and fat-signal separation in liver-fat quantification: comparison to liver biopsy. *Invest Radiol* 2010; 45:465–470.
13. Reeder SB, Pineda AR, Wen Z, et al.: Iterative decomposition of water and fat with echo asymmetry and least-squares estimation (IDEAL): Application with fast spin-echo imaging. *Magn Reson Med* 2005; 54:636–644.
14. Reeder SB, Hu HH, Sirlin CB: Proton Density Fat-Fraction: A Standardized MR-Based Biomarker of Tissue Fat Concentration. *J Magn Reson Imaging* 2012; 36:1011.
15. Yu H, McKenzie CA, Shimakawa A, et al.: Multiecho reconstruction for simultaneous water-fat decomposition and T2* estimation. *Journal of Magnetic Resonance Imaging* 2007; 26:1153–1161.
16. Bley TA, Wieben O, François CJ, Brittain JH, Reeder SB: Fat and water magnetic resonance imaging. *Journal of Magnetic Resonance Imaging* 2010; 31:4–18.
17. Wu EX, Kim D, Tosti CL, et al.: Magnetic resonance assessment of iron overload by separate measurement of tissue ferritin and hemosiderin iron. *Ann N Y Acad Sci* 2010; 1202:115–122.

18. Starekova J, Zhao R, Colgan TJ, et al.: Improved Free-breathing Liver Fat and Iron Quantification using a 2D Chemical Shift Encoded MRI with Flip-Angle Modulation and Motion Corrected Averaging. *Eur Radiol* 2022; 32:5458.
19. Heidenreich JF, Tang J, Tamada D, et al.: Motion-Insensitive Flip Angle Modulated Liver Proton Density Fat-Fraction and R2* Mapping During Free-Breathing MRI in a Clinical Setting. *J Magn Reson Imaging* 2025; 62:1452–1463.
20. Morani AC, Elsayes KM, Liu PS, et al.: Abdominal applications of diffusion-weighted magnetic resonance imaging: Where do we stand. *World J Radiol* 2013; 5:68.
21. Taouli B, Beer AJ, Chenevert T, et al.: Diffusion-weighted imaging outside the brain: Consensus statement from an ISMRM-sponsored workshop. *J Magn Reson Imaging* 2016; 44:521–540.
22. Bharwani N, Koh DM: Diffusion-weighted imaging of the liver: an update. *Cancer Imaging* 2013; 13:171.
23. Obara M, Kwon J, Yoneyama M, Ueda Y, Van Cauteren M: Technical Advancements in Abdominal Diffusion-weighted Imaging. *Magnetic Resonance in Medical Sciences* 2023; 22:191.
24. Herek D, Karabulut N, Kocyıgıt A, Yagcı AB: Evaluation of Free Breathing Versus Breath Hold Diffusion Weighted Imaging in Terms Apparent Diffusion Coefficient (ADC) and Signal-to-Noise Ratio (SNR) Values for Solid Abdominal Organs. *Pol J Radiol* 2016; 81:502.
25. Kwee TC, Takahara T, Koh DM, Nievelstein RAJ, Luijten PR: Comparison and reproducibility of ADC measurements in breathhold, respiratory triggered, and free-breathing diffusion-weighted MR imaging of the liver. *Journal of Magnetic Resonance Imaging* 2008; 28:1141–1148.
26. Chen X, Qin L, Pan D, et al.: Liver Diffusion-weighted MR Imaging: Reproducibility Comparison of ADC Measurements Obtained with Multiple Breath-hold, Free-breathing, Respiratory-triggered, and Navigator-triggered Techniques. *https://doi.org/101148/radiol13131572* 2014; 271:113–125.
27. Hernando D, Liang ZP, Kellman P: Chemical Shift-Based Water/Fat Separation: A Comparison of Signal Models. *Magnetic resonance in medicine : official journal of the Society of Magnetic Resonance in Medicine / Society of Magnetic Resonance in Medicine* 2010; 64:811.
28. Eggers H, Börnert P: Chemical shift encoding-based water-fat separation methods. *J Magn Reson Imaging* 2014; 40:251–268.
29. Anderson AW, Gore JC: Analysis and correction of motion artifacts in diffusion weighted imaging. *Magn Reson Med* 1994; 32:379–387.
30. Aliotta E, Wu HH, Ennis DB: Convex optimized diffusion encoding (CODE) gradient waveforms for minimum echo time and bulk motion-compensated diffusion-weighted MRI. *Magn Reson Med* 2017; 77:717–729.
31. Peña-Nogales Ó, Zhang Y, Wang X, et al.: Optimized Diffusion-Weighting Gradient Waveform Design (ODGD) formulation for motion compensation and concomitant gradient nulling. *Magn Reson Med* 2019; 81:989–1003.
32. Zhang Y, Peña-Nogales Ó, Holmes JH, Hernando D: Motion-robust and blood-suppressed M1-optimized diffusion MR imaging of the liver. *Magn Reson Med* 2019; 82:302–311.

33. Yoon JH, Nickel MD, Peeters JM, Lee JM: Rapid Imaging: Recent Advances in Abdominal MRI for Reducing Acquisition Time and Its Clinical Applications. *Korean J Radiol* 2019; 20:1597.
34. Feng L, Chandarana H: Accelerated Abdominal MRI: A Review of Current Methods and Applications. *Journal of Magnetic Resonance Imaging* 2025; 62:654–672.
35. Chandarana H, Doshi AM, Shanbhogue A, et al.: Three-dimensional MR Cholangiopancreatography in a Breath Hold with Sparsity-based Reconstruction of Highly Undersampled Data. *Radiology* 2016; 280:585.
36. Song JS, Kim SH, Kuehn B, Paek MY: Optimized Breath-Hold Compressed-Sensing 3D MR Cholangiopancreatography at 3T: Image Quality Analysis and Clinical Feasibility Assessment. *Diagnostics* 2020; 10:376.
37. Zhu L, Wu X, Sun Z, et al.: Compressed-Sensing Accelerated 3-Dimensional Magnetic Resonance Cholangiopancreatography: Application in Suspected Pancreatic Diseases. *Invest Radiol* 2018; 53:150–157.
38. Ehman R, McNamara M, Pallack M, et al.: Magnetic resonance imaging with respiratory gating: techniques and advantages. *https://www.ajronline.org/* 2012; 143:1175–1182.
39. Hope TA, Verdin EF, Bergsland EK, Ohliger MA, Corvera CU, Nakakura EK: Correcting for respiratory motion in liver PET/MRI: preliminary evaluation of the utility of bellows and navigated hepatobiliary phase imaging. *EJNMMI Physics 2015 2:1* 2015; 2:21-.
40. Kromrey ML, Funayama S, Tamada D, et al.: Clinical Evaluation of Respiratory-triggered 3D MRCP with Navigator Echoes Compared to Breath-hold Acquisition Using Compressed Sensing and/or Parallel Imaging. *Magnetic Resonance in Medical Sciences* 2019; 19:318.
41. Jin D, Li X, Qian Y, et al.: Modified respiratory-triggered SPACE sequences for magnetic resonance cholangiopancreatography. *Eur J Radiol Open* 2024; 12.
42. Dixon WT, Brummer ME, Malko JA: Acquisition order and motional artifact reduction in spin warp images. *Magn Reson Med* 1988; 6:74–83.
43. Block KT, Chandarana H, Milla S, et al.: Towards Routine Clinical Use of Radial Stack-of-Stars 3D Gradient-Echo Sequences for Reducing Motion Sensitivity. *Journal of the Korean Society of Magnetic Resonance in Medicine* 2014; 18:87.
44. Feng L: Golden-Angle Radial MRI: Basics, Advances, and Applications. *Journal of Magnetic Resonance Imaging* 2022.
45. Chandarana H, Block TK, Rosenkrantz AB, et al.: Free-breathing radial 3D fat-suppressed T1-weighted gradient echo sequence: A viable alternative for contrast-enhanced liver imaging in patients unable to suspend respiration. *Invest Radiol* 2011; 46:648–653.
46. Pipe JG: Motion Correction With PROPELLER MRI: Application to Head Motion and Free-Breathing Cardiac Imaging. *Magn Reson Med* 1999; 42:963.
47. Maintz JBA, Viergever MA: A survey of medical image registration. *Med Image Anal* 1998; 2:1–36.
48. Zitová B, Flusser J: Image registration methods: a survey. *Image Vis Comput* 2003; 21:977–1000.
49. Klein A, Andersson J, Ardekani BA, et al.: Evaluation of 14 nonlinear deformation algorithms applied to human brain MRI registration. *Neuroimage* 2009; 46:786–802.

50. Sotiras A, Davatzikos C, Paragios N: Deformable Medical Image Registration: A Survey. *IEEE Trans Med Imaging* 2013; 32:1153.
51. Schnabel JA, Heinrich MP, Papież BW, Brady SJM: Advances and challenges in deformable image registration: From image fusion to complex motion modelling. *Med Image Anal* 2016; 33:145–148.
52. Xu Z, Lee CP, Heinrich MP, et al.: Evaluation of Six Registration Methods for the Human Abdomen on Clinically Acquired CT. *IEEE Trans Biomed Eng* 2016; 63:1563.
53. Son JS, Park HS, Park S, et al.: Motion-Corrected versus Conventional Diffusion-Weighted Magnetic Resonance Imaging of the Liver Using Non-Rigid Registration. *Diagnostics 2023, Vol 13,* 2023; 13.
54. Pham D, Kron T, Foroudi F, Schneider M, Siva S: A review of kidney motion under free, deep and forced-shallow breathing conditions: implications for stereotactic ablative body radiotherapy treatment. *Technol Cancer Res Treat* 2014; 13:315–323.
55. McTavish S, Van AT, Peeters JM, et al.: Motion compensated renal diffusion weighted imaging. *Magn Reson Med* 2023; 89:144–160.
56. Coll-Font J, Afacan O, Hoge S, et al.: Retrospective distortion and motion correction for free-breathing DW-MRI of the kidneys using dual echo EPI and slice-to-volume registration. *J Magn Reson Imaging* 2020; 53:10.1002/jmri.27473.
57. Ariyurek C, Wallace TE, Kober T, Kurugol S, Afacan O: Prospective motion correction in kidney MRI using FID navigators. *Magn Reson Med* 2023; 89:276–285.
58. Lenz GW, Haacke EM, White RD: Retrospective cardiac gating: a review of technical aspects and future directions. *Magn Reson Imaging* 1989; 7:445–455.
59. Feng L, Benkert T, Block KT, Sodickson DK, Otazo R, Chandarana H: Compressed sensing for body MRI. *Journal of Magnetic Resonance Imaging* 2017; 45:966–987.
60. Grimm R, Fürst S, Dregely I, et al.: Self-gated radial MRI for respiratory motion compensation on hybrid PET/MR systems. *Med Image Comput Comput Assist Interv* 2013; 16(Pt 3):17–24.
61. Feng L, Grimm R, Block KT obias, et al.: Golden-angle radial sparse parallel MRI: combination of compressed sensing, parallel imaging, and golden-angle radial sampling for fast and flexible dynamic volumetric MRI. *Magn Reson Med* 2014; 72:707–717.
62. Piccini D, Littmann A, Nielles-Vallespin S, Zenge MO: Respiratory self-navigation for whole-heart bright-blood coronary MRI: Methods for robust isolation and automatic segmentation of the blood pool. *Magn Reson Med* 2012; 68:571–579.
63. Bonanno G, Puy G, Wiaux Y, Van Heeswijk RB, Piccini D, Stuber M: Self-navigation with compressed sensing for 2D translational motion correction in free-breathing coronary MRI: A feasibility study. *PLoS One* 2014; 9.
64. Pang J, Sharif B, Fan Z, et al.: ECG and navigator-free four-dimensional whole-heart coronary MRA for simultaneous visualization of cardiac anatomy and function. *Magn Reson Med* 2014; 72:1208–1217.
65. Pang J, Bhat H, Sharif B, et al.: Whole-heart coronary MRA with 100% respiratory gating efficiency: Self-navigated three-dimensional retrospective image-based motion correction (TRIM). *Magn Reson Med* 2014; 71:67–74.
66. Rank CM, Heußer T, Buzan MTA, et al.: 4D respiratory motion-compensated image reconstruction of free-breathing radial MR data with very high undersampling. *Magn Reson Med* 2017; 77:1170–1183.

67. Cheng JY, Alley MT, Cunningham CH, Vasanawala SS, Pauly JM, Lustig M: Nonrigid motion correction in 3D using autofocusing with localized linear translations. *Magn Reson Med* 2012; 68:1785–1797.
68. Schmidt JFM, Buehrer M, Boesiger P, Kozerke S: Nonrigid retrospective respiratory motion correction in whole-heart coronary MRA. *Magn Reson Med* 2011; 66:1541–1549.
69. Buerger C, Schaeffter T, King AP: Hierarchical adaptive local affine registration for fast and robust respiratory motion estimation. *Med Image Anal* 2011; 15:551–564.
70. Feng L, Axel L, Chandarana H, Block KT, Sodickson DK, Otazo R: XD-GRASP: Golden-angle radial MRI with reconstruction of extra motion-state dimensions using compressed sensing. *Magn Reson Med* 2016; 75:775–788.
71. Feng L, Delacoste J, Smith D, et al.: Simultaneous Evaluation of Lung Anatomy and Ventilation Using 4D Respiratory-Motion-Resolved Ultrashort Echo Time Sparse MRI. *Journal of Magnetic Resonance Imaging* 2019; 49:411–422.
72. Feng L, Coppo S, Piccini D, et al.: 5D whole-heart sparse MRI. *Magn Reson Med* 2018; 79:826–838.
73. Piccini D, Feng L, Bonanno G, et al.: Four-dimensional respiratory motion-resolved whole heart coronary MR angiography. *Magn Reson Med* 2017; 77:1473–1484.
74. Chandarana H, Feng L, Ream J, et al.: Respiratory motion-resolved compressed sensing reconstruction of free-breathing radial acquisition for dynamic liver magnetic resonance imaging. *Invest Radiol* 2015; 50:749–756.
75. Bruijnen T, Schake T, Akdag O, et al.: Free-breathing motion compensated 4D (3D+respiration) T2-weighted turbo spin-echo MRI for body imaging. *arXiv:220203021 [physics.med-ph]* 2022.
76. Forman C, Piccini D, Grimm R, Hutter J, Hornegger J, Zenge MO: Reduction of respiratory motion artifacts for free-breathing whole-heart coronary MRA by weighted iterative reconstruction. *Magn Reson Med* 2015; 73:1885–1895.
77. Cheng JY, Zhang T, Ruangwattanapaisarn N, et al.: Free-breathing pediatric MRI with nonrigid motion correction and acceleration. *Journal of Magnetic Resonance Imaging* 2015; 42:407–420.
78. Feng L, Huang C, Shanbhogue K, Sodickson DK, Chandarana H, Otazo R: RACER-GRASP: Respiratory-weighted, aortic contrast enhancement-guided and coil-unstreaking golden-angle radial sparse MRI. *Magn Reson Med* 2018; 80:77–89.
79. Jiang W, Ong F, Johnson KM, et al.: Motion robust high resolution 3D free-breathing pulmonary MRI using dynamic 3D image self-navigator. *Magn Reson Med* 2018; 79:2954–2967.
80. Gandhi DB, Higano NS, Hahn AD, et al.: Comparison of Weighting Algorithms to Mitigate Respiratory Motion in Free-Breathing Neonatal Pulmonary Radial UTE-MRI. *Biomed Phys Eng Express* 2024; 10:10.1088/2057-1976/ad3cdd.
81. Chen L, Zeng X, Ji B, et al.: Improving dynamic contrast-enhanced MRI of the lung using motion-weighted sparse reconstruction: Initial experiences in patients. *Magn Reson Imaging* 2020; 68:36–44.
82. Vahle T, Bacher M, Rigie D, et al.: Respiratory Motion Detection and Correction for MR Using the Pilot Tone: Applications for MR and Simultaneous PET/MR Exams. *Invest Radiol* 2020; 55:153.

83. Solomon E, Rigie DS, Vahle T, et al.: Free-breathing radial imaging using a pilot-tone radiofrequency transmitter for detection of respiratory motion. *Magn Reson Med* 2021; 85:2672–2685.

84. Gottwald LM, Blanken CPS, Tourais J, et al.: Retrospective Camera-Based Respiratory Gating in Clinical Whole-Heart 4D Flow MRI. *Journal of Magnetic Resonance Imaging* 2021; 54:440.

85. Feng L, Wen Q, Huang C, Tong A, Liu F, Chandarana H: GRASP-Pro: imProving GRASP DCE-MRI through self-calibrating subspace-modeling and contrast phase automation. *Magn Reson Med* 2020; 83:94–108.

86. Chen J, Huang C, Shanbhogue K, et al.: DCE-MRI of the liver with sub-second temporal resolution using GRASP-Pro with navi-stack-of-stars sampling. *NMR Biomed* 2024.

87. Solomon E, Bae J, Zan E, et al.: GRASP-Pro+: GRASP reconstruction with locally low-rank subspace constraint for DCE-MRI. In *Proc Intl Soc Mag Reson Med 30 (2022) p1603*; .

88. Feng L: 4D Golden-Angle Radial MRI at Subsecond Temporal Resolution. *NMR Biomed* 2022.

89. Feng L: Live-view 4D GRASP MRI: A framework for robust real-time respiratory motion tracking with a sub-second imaging latency. *Magn Reson Med* 2023; 90.

90. Zhang S, Joseph AA, Voit D, et al.: Real-time magnetic resonance imaging of cardiac function and flow-recent progress. *Quant Imaging Med Surg* 2014.

91. Wang X, Uecker M, Feng L: Fast Real-Time Cardiac MRI: a Review of Current Techniques and Future Directions. *Investig Magn Reson Imaging* 2021; 25:252.

92. Frahm J, Voit D, Uecker M: Real-Time Magnetic Resonance Imaging: Radial Gradient-Echo Sequences With Nonlinear Inverse Reconstruction. *Invest Radiol* 2019; 54:757–766.

93. Contijoch F, Rasche V, Seiberlich N, Peters DC: The future of CMR: All-in-one vs. real-time CMR (Part 2). *Journal of Cardiovascular Magnetic Resonance* 2024; 26:100998.

94. Starekova J, Zhao R, Colgan TJ, et al.: Improved free-breathing liver fat and iron quantification using a 2D chemical shift-encoded MRI with flip angle modulation and motion-corrected averaging. *Eur Radiol* 2022; 32:5458–5467.

95. Heidenreich JF, Tang J, Tamada D, et al.: Motion-Insensitive Flip Angle Modulated Liver Proton Density Fat-Fraction and R2* Mapping During Free-Breathing MRI in a Clinical Setting. *J Magn Reson Imaging* 2025; 62:1452–1463.

96. Bellal M, El Fkihi S, Cengiz K, Ivkovic N: A Comprehensive Survey on Deep Learning in Abdominal Imaging: Datasets, Techniques, and Performance Metrics. *IEEE Access* 2025; 13:79894–79914.

97. Kiryu S, Akai H, Yasaka K, et al.: Clinical Impact of Deep Learning Reconstruction in MRI. *Radiographics* 2023; 43.

98. Küstner T, Armanious K, Yang J, Yang B, Schick F, Gatidis S: Retrospective correction of motion-affected MR images using deep learning frameworks. *Magn Reson Med* 2019; 82:1527–1540.

99. Tamada D, Kromrey ML, Ichikawa S, Onishi H, Motosugi U: Motion Artifact Reduction Using a Convolutional Neural Network for Dynamic Contrast Enhanced MR Imaging of the Liver. *Magn Reson Med Sci* 2020; 19:64–76.

100. Cui L, Song Y, Wang Y, et al.: Motion artifact reduction for magnetic resonance imaging with deep learning and k-space analysis. *PLoS One* 2023; 18:e0278668.

101. Terpstra ML, Maspero M, D'Agata F, et al.: Deep learning-based image reconstruction and motion estimation from undersampled radial k-space for real-time MRI-guided radiotherapy. *Phys Med Biol* 2020; 65:155015.

102. Miller Z, Johnson KM: Motion compensated self supervised deep learning for highly accelerated 3D ultrashort Echo time pulmonary MRI. *Magn Reson Med* 2023; 89:2361–2375.

103. Spieker V, Eichhorn H, Hammernik K, et al.: Deep Learning for Retrospective Motion Correction in MRI: A Comprehensive Review. *IEEE Trans Med Imaging* 2024; 43:846–859.

104. Ghoul A, Pan J, Lingg A, et al.: Attention-Aware Non-Rigid Image Registration for Accelerated MR Imaging. *IEEE Trans Med Imaging* 2024; 43:3013–3026.

105. Pan J, Huang W, Rueckert D, Kustner T, Hammernik K: Motion-Compensated MR CINE Reconstruction With Reconstruction-Driven Motion Estimation. *IEEE Trans Med Imaging* 2024; 43:2420–2433.

106. Feng J, Chen J, Zhang Y, Feng L, Liang D, Wei H: Calibration-free DCE-MRI with Sub-second Temporal Resolution Using Interpretable Implicit Neural Representation. In *2025 ISMRM Annual Meeting Proceedings, Honolulu, USA p123*; .

107. Gadjimuradov F, Benkert T, Nickel MD, Führes T, Saake M, Maier A: Deep learning–guided weighted averaging for signal dropout compensation in DWI of the liver. *Magn Reson Med* 2022; 88:2679–2693.

108. Lehtinen J, Munkberg J, Hasselgren J, et al.: Noise2Noise: Learning Image Restoration without Clean Data. *35th International Conference on Machine Learning, ICML 2018* 2018; 7:4620–4631.

109. Janjušević N, Chen J, Ginocchio L, et al.: Self-Supervised Noise Adaptive MRI Denoising via Repetition to Repetition (Rep2Rep) Learning. *Magn Reson Med* 2026; 95:1619–1633.

110. Pfaff L, Darwish O, Wagner F, et al.: Enhancing diffusion-weighted prostate MRI through self-supervised denoising and evaluation. *Scientific Reports 2024 14:1* 2024; 14:24292-.

111. Chen J, Pei H, Maier C, et al.: Self-Supervised Joint Reconstruction and Denoising of T2-Weighted PROPELLER MRI of the Lung at 0.55T. *Magn Reson Med* 2025.